\documentclass[a4paper,11pt]{article}
\pdfoutput=1 

\usepackage{jcappub}
\usepackage[T1]{fontenc} 
\usepackage{comment}
\usepackage{empheq}
\usepackage{amsmath,amssymb,amsbsy}
\usepackage{bbold}
\usepackage{slashed}
\usepackage{subcaption}
\usepackage{tikz}
\usepackage{soul}
\usepackage{physics}
\usepackage{multirow}
\usepackage{siunitx}
\usepackage{tcolorbox}
\usepackage{mathtools}

\newtcbox{\eqbox}{colback=gray!10, colframe=gray!40, boxrule=0.4pt, arc=2pt,
                  boxsep=0pt, left=4pt, right=4pt, top=3pt, bottom=3pt,
                  nobeforeafter, tcbox raise base}

\newcommand{\bfp}[0]{\mathbf{p}}

\title{Real Intermediate States in Dark Matter Freeze-In from First Principles: Single Mediator}

\author[a]{Jannis Brirup,}
\author[a]{Adrian Finke,}
\author[a]{Michael Klasen}
\author[b]{and Luca Paolo Wiggering}

\affiliation[a]{Institut für Theoretische Physik, Universität Münster, Wilhelm-Klemm-Straße 9, 48149 Münster, Germany}
\affiliation[b]{Centre for Cosmology, Particle Physics and Phenomenology (CP3), Université Catholique de Louvain, Chemin du Cyclotron 2, 1348 Louvain-la-Neuve, Belgium}

\emailAdd{jannis.brirup@uni-muenster.de}
\emailAdd{adrian.finke@uni-muenster.de}
\emailAdd{michael.klasen@uni-muenster.de}
\emailAdd{luca.wiggering@uclouvain.be}

\abstract{
We investigate the treatment of real intermediate states (RIS) in freeze-in production of $s$-channel Dark Matter (DM) with unstable mediators, as such models feature a double-counting issue when decay and annihilation collision terms are simply added within the semi-classical Boltzmann approach and a finite width is used for the regularization of the resonance.
While several ad-hoc RIS-subtraction schemes are available to mitigate this problem, a first-principles treatment in this context is so far missing.
We therefore start from the Closed-Time-Path (CTP) formulation of non-equilibrium Quantum Field Theory and present a CTP-based description of $s$-channel freeze-in pair production of a single scalar, fermion or vector DM species. In addition to being intrinsically free of non-unitary overcounting, the CTP ansatz also has the advantage of systematically capturing thermal corrections known to be important for DM freeze-in. Separate evolution equations assuming a single scalar or vector mediator in thermal equilibrium with the surrounding bath are presented. 
The exact propagators of the unstable mediators are approximated by resumming one particle irreducible (1PI) self-energy diagrams generated by bath particles 
and are evaluated with their full mass and momentum dependence in contrast to the commonly adopted Hard Thermal Loop limit.
The CTP relic densities are compared numerically with those obtained using the most common RIS-subtraction schemes for two example $s$-channel freeze-in models, demonstrating the phenomenological relevance of a consistent RIS treatment.
}

\begin{document}
\hfill MS-TP-26-17\\[5mm]
\maketitle
\flushbottom

\section{Introduction}

Despite the compelling evidence for dark matter (DM) on galactic~\cite{Rubin:1980zd}, cluster~\cite{Clowe:2006eq} and cosmological scales~\cite{Planck:2018vyg}, its microscopic nature remains unknown~\cite{Cirelli:2024ssz}. Even though the thermal freeze-out of weakly interacting massive particles (WIMPs) from the primordial Standard Model plasma~\cite{Lee:1977ua,Gondolo:1990dk,Jungman:1995df} ``miraculously'' reproduces the present observationally determined DM abundance $\Omega_{\rm DM} h^2 \simeq 0.120$~\cite{Planck:2018vyg}, the lack of clear evidence for WIMPs at colliders~\cite{CMS:2024zqs}, in direct~\cite{LZ:2024zvo} and indirect detection~\cite{MAGIC:2016xys} experiments has significantly increased the interest in alternative DM candidates with different interactions strengths and production mechanisms. One well-motivated alternative possibility are feebly interacting massive particles (FIMPs)~\cite{Bernal:2017kxu} which, in contrast to WIMPs, have a negligible abundance after reheating due to their tiny interaction strengths and which are instead produced subsequently via freeze-in~\cite{Dodelson:1993je,Hall:2009bx} through decays and annihilations of thermal bath particles. However, in the semi-classical Boltzmann approach~\cite{bernstein_1988} widely adopted for the computation of the freeze-in abundance, the DM production via $s$-channel annihilation with a Breit--Wigner regulated resonance leads to a double-counting inconsistency in the parameter region where mediator decays into DM are kinematically allowed, if the annihilation and decay collision terms are naively added. To circumvent this problem, several ad-hoc schemes have been proposed. They either 1) supply a subtraction term to somehow remove the real intermediate state (RIS) from the Breit--Wigner propagator~\cite{Kolb:1979qa, Kolb:1979ui, Cline:1993bd,Nardi:2007jp,Ala-Mattinen:2019mpa, Ala-Mattinen:2023rbm} or 2) modify the width in the propagator denominator to render the decay collision term obsolete~\cite{Belanger:2018ccd,Bringmann_2022}. In addition, the majority of these prescriptions only applies to the modulus squared of a single Breit--Wigner propagator, and the generalization to interference terms with only one or two different resonant propagators is not straightforward. At the same time, since DM is still relativistic during its production, it is known that the Fermi--Dirac or Bose--Einstein nature of the bath particles has to be taken into account~\cite{Drewes:2015eoa,Lebedev:2019ton,Belanger:2018ccd,Belanger:2020npe,Bringmann_2022} in addition to in-medium modifications of the spectrum of propagating states and their dispersion relations~\cite{Drewes:2015eoa,Biondini:2020ric,Bringmann_2022,Becker:2023vwd}. While quantum-statistical effects are still captured in the Boltzmann collision term and can lead, for scenarios where the mediator can be produced on shell, to a deviation of the DM abundance at the level of $\sim\SI{10}{\percent}$~\cite{Belanger:2018ccd} compared to the usage of Maxwell--Boltzmann statistics, this semi-classical approach does not provide a consistent framework for finite-temperature corrections, and the double-counting issue remains unresolved as well. We therefore adopt a first-principles path and derive evolution equations for the DM number density starting from the Keldysh--Schwinger~\cite{10.1063/1.1703727-julianschwinger,Keldysh:1964ud} formulation of non-equilibrium Quantum Field Theory (QFT)~\cite{Danielewicz:1982kk,Chou:1984es,Calzetta:1986cq,Rammer:1986zz,Botermans:1990qi,Berges_2004}. Besides being intrinsically free of the double-counting redundancy, it also has the advantage of consistently capturing in-medium effects. For the same reasons, the Keldysh--Schwinger formalism has already been extensively applied to baryogenesis \cite{Riotto:1995hh,Buchmuller:2000nd,Giudice:2003jh,Prokopec:2003pj,Prokopec:2004ic,Garny:2009rv,Beneke:2010wd,Fidler:2011yq,Salvio:2011sf,Garny:2011hg,Frossard:2012pc,BhupalDev:2014oar,Drewes:2016gmt,Garbrecht:2018mrp,Kainulainen:2021oqs} and 
is gaining popularity for DM freeze-in~\cite{Drewes:2015eoa,Ai:2023qnr,Abada:2023mib,Becker:2023vwd,Becker:2025lkc}. {In particular, Ref.~\cite{Becker:2023vwd} computed the freeze-in abundance of scalar DM produced via a gauge-charged fermionic mediator with 1PI-resummed propagators beyond the Hard Thermal Loop (HTL) approximation, keeping the full dependence on the vacuum mass scale. There, the mediator never appears as an $s$-channel resonance and the RIS problem does not arise. The present work is complementary: we resum a bosonic (scalar or vector) $s$-channel mediator and focus on the double counting between its decay and resonant annihilations.}

Let us note that, in the context of DM freeze-out, RIS subtraction has also been addressed from first principles within the Boltzmann formalism~\cite{Matak:2022qwc,Matak:2023zox} by including all kinematically allowed holomorphic cuts of the forward-scattering diagram for DM annihilation in vacuum QFT with the result of a finite fixed-order annihilation cross section. Ref.~\cite{Laine:2022ner} on the other hand performed a fixed-order next-to-leading order computation of DM annihilation within the Matsubara imaginary-time formalism~\cite{Matsubara:1955ws,LeBellac_1996,Laine:2016hma}.
 The need for a careful treatment of mass singularities from internal propagators becoming on shell is also not restricted to $s$-channel processes. For example, Refs.~\cite{Grzadkowski:2021kgi,Iglicki:2022jjf} discuss the regularization of $t$-channel singularities in early Universe computations by introducing a thermal width for the unstable particle, while Ref.~\cite{Beocanin:2026cjt} showed similarly to~\cite{Matak:2022qwc,Matak:2023zox} that also in vacuum at fixed order a finite observable may be constructed by summing all cuts of a single forward scattering diagram.

The paper is organized as follows:
In Sec.~\ref{sec:Freeze-in in the Boltzmann Formalism}, we introduce the standard Boltzmann description of DM freeze-in production and discuss several RIS-subtraction schemes available in the literature to avoid the overcounting problem. Then, in Sec.~\ref{sec:Freeze-in in the CTP Formalism}, we start from the Kadanoff--Baym equations to derive generic Closed-Time-Path (CTP) evolution equations for the freeze-in number density of real scalar, fermion and vector DM from first principles.
In Sec.~\ref{sec:CTP-Resummed}, we then focus on $s$-channel production and provide explicit expressions for the DM production rate in terms of the 1PI-resummed mediator spectral density, distinguishing between a scalar and a vector mediator. We show how this production rate can be understood as an $s$-channel Boltzmann collision term, but with a modified Breit--Wigner propagator. In Sec.~\ref{sec:order-by-order}, to establish the correspondence to the Boltzmann formalism, we analyze the CTP DM production rate order-by-order in mediator self-energy insertions. The contribution with a free mediator propagator corresponds exactly to the Boltzmann mediator decay collision term, while a single mediator self-energy insertion gives rise to $s$-channel production processes, such as $2\to2$ annihilation. Unlike in the standard finite mediator-width approach, no double counting of the intermediate state occurs in the fixed-order computation in agreement with the findings of Refs.~\cite{Matak:2022qwc,Matak:2023zox}.
In Sec.~\ref{sec: Numerical Examples}, we consider two example freeze-in DM models: one with a scalar mediator and one with a vector mediator. This is to illustrate the application of our CTP-based evolution equations and to show how they numerically compare in terms of the resulting DM abundance today with the other ad-hoc RIS-subtraction schemes around the resonant parameter regions. Conclusions are drawn in Sec.~\ref{sec:Conclusion}.

\section{Freeze-In in the Boltzmann Formalism and the RIS Problem}
\label{sec:Freeze-in in the Boltzmann Formalism}

We start with a very brief summary of the standard description of DM freeze-in production using the Boltzmann formalism and provide a list of ad-hoc schemes to address the associated double-counting issue of real intermediate states. 

\subsection{Boltzmann Equation}

 In the Boltzmann approach, the evolution of the DM number density $n_{\rm DM}$ in an expanding FLRW spacetime with scale factor $a(t)$ is governed by the differential equation 
\begin{equation}
    \dot{n}_{\rm DM} +3 H  {n}_{\rm DM} = \mathcal{C}[f_{\rm DM}]\, ,
    \label{Eq: BoltzmannEq}
\end{equation}
where a dot denotes the derivative with respect to cosmic time $t$, $H=\dot{a}/a$ is the Hubble rate, and $\mathcal{C}[f_{\rm DM}]$ is the collision term summing all number-changing reactions that impact the DM phase space density $f_{\rm DM}$. For the case of a $2\to 2$ process, most relevant to the subsequent discussion, the collision term in Eq.~\eqref{Eq: BoltzmannEq} is defined such that it counts a reaction twice if two identical particles are produced and such that $n_{\rm DM}$ contains in the case of non self-conjugated DM both, the particle and anti-particle number densities.

\paragraph{Mediator Decay} For a mediator $Y$ decaying into two generic particles $A$ and $B$ that could come from the thermal bath or constitute the DM, the $1\to 2$ decay collision term 
\begin{equation}
    \mathcal{C}_{Y \to A,B} = \int \dd{\Pi_Y}\dd{\Pi_A}\dd{\Pi_B}(2\pi)^4 \delta^{(4)}(p_Y - p_A - p_B) \, |\mathcal{M}|^2_{Y\to A, B}\,  f_Y \, \bar{f}_A \, \bar{f}_B \, 
    \label{eq:C:Y_to_ab}
\end{equation}
integrates the squared and summed vacuum decay matrix element $|\mathcal{M}|^2_{Y\to A, B}$ over all kinematically allowed momentum configurations via the Lorentz invariant phase space element 
\begin{equation}
    \dd{\Pi} = \frac{\dd[3]{\bfp}}{(2\pi)^3 2 E}
\end{equation}
with energy $E$. The final-state phase space densities $f_{A,B}$ enter through the quantum statistical factors $\bar{f}_i = 1 + \eta_i f_i$ describing Bose enhancement $(\eta_i=+1)$ for bosons and Pauli blocking $(\eta_i=-1)$ for fermions. In equilibrium, for a vanishing chemical potential, the phase space densities take the well-known form $f^{\rm eq}_i(E) = 1/({e^{\beta E } - \eta_i})$ with the inverse temperature $\beta = 1/T$.
Due to Lorentz invariance, $|\mathcal{M}|^2_{Y\to A, B}$ features at leading order no angular dependence, so that the final-state integration factorizes in that case into the vacuum decay rate $\Gamma^{\rm vac}_{Y\to A,B}$ times an in-medium correction 
\begin{equation}
    \bar{G}_{AB}(\bfp_Y, s) =   \int_{-1} ^{1} \frac{\dd{\!\cos\theta^{\rm CM}}}{2} \, [1+ \eta_A\, f_A(E_A)][1+\eta_B \,f_B(E_B)],
    \label{eq:Gab}
\end{equation}
accounting for Bose enhancement or Fermi blocking. As a result, the decay collision term simplifies to the product
\begin{equation}
    \mathcal{C}_{Y\to A,B} =  \Gamma^{\rm vac}_{Y\to A,B}\int \frac{\dd[3]{\mathbf{p}_Y}}{(2\pi)^3 }\,  f_Y(E_Y) \,  \bar{G}_{AB}(\mathbf{p}_Y,m_Y^2)\, .
\end{equation}
The integration in Eq.~\eqref{eq:Gab} is performed over the cosine of the center-of-mass (CM) angle $\theta^{\rm CM}$ between the three-momentum of $A$ and the Lorentz boost from the CM frame of the decay products to the cosmic frame. It therefore interpolates linearly between the minimum and maximum energy of $A$ and $B$ in the cosmic rest frame via
\begin{align}
  E_{A/B} (\cos\theta^{\rm CM})   = \frac{1}{\sqrt{s} } \left|{E_Y}  E_{A/B}^{\rm CM} \pm  |\bfp_Y| |\bfp^{\rm CM}| \cos\theta^{\rm CM} \right| \, .
    \label{eq:Eab_limits}
\end{align}
The corresponding CM energies and momenta are entirely fixed by $s=p_Y^2$ and the masses $m_A$ and $m_B$. They read
\begin{subequations}
    \begin{align}
    &|\bfp^{\rm CM}| =\sqrt{(s + m^2_A - m^2_B)^2 - 4 sm^2_A}\Big/(2 \sqrt{s}) \,  , \\
    &E^{\rm CM}_{A/B} = (s \pm m^2_A \mp m^2_B)/(2\sqrt{s})\, .
\end{align}
\end{subequations}
If $A$ and $B$ are furthermore in full thermal equilibrium, the integral in Eq.~\eqref{eq:Gab} can be performed analytically with the result~\cite{Belanger:2018ccd, Bringmann_2022}\footnote{Ref.~\cite{Bringmann_2022} uses the Lorentz factor $\gamma=p_Y^0/\sqrt{s}$ to parametrize the function $\bar{G}_{AB}$ instead of the three-momentum $\mathbf{p}_Y$.}
\begin{align}
     \bar{G}_{AB}(\bfp_Y, s)  
     &= \frac{1}{1 - \eta_A\eta_B e^{-\beta E_Y}} \frac{\sqrt{s}} {2\beta |\bfp_Y||\bfp^{\rm CM}| } \, \log\!\left(\frac{ e^{\beta E_A(1)} - \eta_A }{e^{\beta E_A(-1) } - \eta_A}\, \frac{ 1 - \eta_B e^{-\beta E_B(-1) } }{1 - \eta_B e^{-\beta E_B(1) }}   \right) \, .
\end{align}

\paragraph{Annihilation of Bath Particles} Besides mediator decays, DM may also be produced through annihilation processes from bath particles $A$ and $B$. The corresponding $2\to 2$ collision term reads
\begin{multline}
    \mathcal{C}_{A,B \to \text{DM},\overline{\text{DM}}} = \int  \dd{\Pi}_{A} \dd{\Pi}_{B} \dd{\Pi}_{\text{DM}} \dd{\Pi}_{\overline{\text{DM}}} \\ \times  (2\pi)^4 \delta^{(4)}( p_A + p_B - p_\text{DM} - p_{\overline{\text{DM}}})  \, |\mathcal{M}|^2_{A, B \to \text{DM}, \overline{\text{DM}}} \, {f}_A {f}_B\bar{f}_\text{DM} \bar{f}_{\overline{\text{DM}}} \, .
    \label{eq:Cann}
\end{multline}
Since we consider only a single DM species in this work, ``$\overline{\text{DM}}$'' has the meaning of the $CP$-conjugate. In the standard freeze-in framework, the initial DM abundance is taken to be negligible, $f_{\mathrm{DM}} \ll 1$ or, equivalently, $\bar{f}_{\mathrm{DM}}\approx 1$. The important point is that for a resonant annihilation process, the mediator propagator in the vacuum matrix element in Eq.~\eqref{eq:Cann} is regularized through its Breit--Wigner form
\begin{align}
\Delta_{\rm BW}(s) = \frac{1}{s-m_Y^2+i m_Y\Gamma^{\rm vac}_Y},
\qquad 
|\Delta_{\rm BW}(s)|^2 = \frac{1}{(s-m^2_Y)^2+(m_Y\Gamma^{\rm vac}_Y)^2}
\label{eq:BW_prop}
\end{align}
with the mediator mass $m_Y$ and its total vacuum decay width $\Gamma^{\rm vac}_Y$. In the narrow-width limit and without final-state blocking and enhancement, Eq.~\eqref{eq:Cann} reproduces, under the additional assumption of Maxwell-Boltzmann statistics, the decay term in Eq.~\eqref{eq:C:Y_to_ab}~\cite{Ala-Mattinen:2023rbm}, which is the origin of the double counting addressed by the schemes summarized next. 

\subsection{Overview of RIS-Subtraction Schemes}
\label{sec: The Real Intermediate State Problem}

The two central RIS--subtraction approaches are either to 1) somehow remove the on-shell part from the Breit--Wigner propagator, Eq.~\eqref{eq:BW_prop}, or to 2) redefine the width in the propagator denominator such that it automatically captures mediator decays, so that the associated collision term can be dropped from the Boltzmann equation. In the following, we list multiple schemes belonging to both categories that have been applied in the literature. For clarity, most mediator subscripts are dropped.

\begin{itemize}
\item \textit{Delta-function subtraction (SDF):}
The delta-function subtraction scheme \cite{Kolb:1979qa,Kolb:1979ui,Cline:1993bd} is defined at the level of the squared Breit--Wigner propagator by removing its narrow-width part,
\begin{align}
|\Delta^\text{SDF}_\text{BW}(s)|^2\Big|_\text{off}=|\Delta_\text{BW}(s)|^2 - \frac{\pi}{m\Gamma}\delta(s-m^2).
\end{align}

\item \textit{Real-part subtraction:}
In this prescription~\cite{Ala-Mattinen:2019mpa,Cline:1993bd,Cline:2017qpe} the real part of the Breit--Wigner propagator is identified with the off-shell contribution,
\begin{align}\label{eq: real-subtraction}
    \Delta^\text{real}_\text{BW}(s)\Big|_\text{off}=    \mathrm{Re}\,\Delta_{\rm BW}(s) = \frac{s-m^2}{(s-m^2)^2+(m\Gamma)^2}\, .
\end{align}
At the level of the squared off-shell propagator one then obtains
\begin{align}
    |\Delta^\text{real}_\text{BW}(s)|^2\Big|_\text{off} = \frac{(s-m^2)^2}{\left[(s-m^2)^2+(m\Gamma)^2\right]^2}\, .
\end{align}
The contribution excluded by this prescription can be made explicit by considering
\begin{align}
  \nonumber  \frac{1}{(s-m^2)^2+(m\Gamma)^2} -\frac{(s-m^2)^2}{[(s-m^2)^2+(m\Gamma)^2]^2}
  &  = \frac{(m\Gamma)^2}{[(s-m^2)^2+(m\Gamma)^2]^2}\\&\xrightarrow[m\Gamma\to0]{}\; \frac{\pi}{2m\Gamma}\delta(s-m^2), 
  \label{eq: realproof}
\end{align}
where the delta-function representation $\delta(x) = \lim_{\epsilon \to 0^+} \frac{2}{\pi} \frac{\epsilon^3}{(x^2 + \epsilon^2)^2 }$ has been used.
Thus, in the zero-width limit, the squared real-part prescription subtracts only half of the narrow-width contribution.

\item \textit{Principal-value subtraction (PVS):}
For a squared propagator, Ref.~\cite{Cline:1993bd} instead defines the off-shell part of the squared propagator in the principal-value subtraction scheme as
\begin{align}
  |\Delta^\text{PVS}_\text{BW}(s)|^2\Big|_\text{off} = \frac{(s-m^2)^2-(m\Gamma)^2}{\left[(s-m^2)^2+(m\Gamma)^2\right]^2}\, ,
  \label{eq:PVS}
\end{align}
since in the zero-width limit this expression corresponds to the principal-value distribution associated with the squared propagator.\footnote{More precisely, this is the Hadamard finite part of $(s-m^2)^{-2}$; we follow the common terminology of the RIS-subtraction literature and refer to it as a principal-value prescription.} 
At the level of the squared propagator, the PVS and SDF schemes are equivalent up to corrections of order $\Gamma^2$ \cite{Cline:1993bd, Ala-Mattinen:2023rbm}. 

\item \textit{Cut-scheme:}
The cut-scheme~\cite{DeRomeri:2020wng,Ala-Mattinen:2023rbm} discards the resonant region of the $2\to 2$ process through cuts on the phase space integral, i.e., those momentum modes within distance $\delta$ of the singularity are associated with the on-shell part of the propagator via some selection function $\Theta(x,\delta)$ that can be as simple as the rectangular function $\Theta(x,\delta) = \theta(x+\delta)\theta(\delta-x)$ with the Heaviside function $\theta$. The corresponding off-shell region of the Breit--Wigner propagator is in this scheme therefore defined as
\begin{equation}
    \Delta^{\rm Cut}_{\rm BW}(s)\Big|_\text{off} = [1-\Theta(s-m^2,\delta)]\, \Delta_{\rm BW}(s) \, .
\end{equation}
In addition, to account for the fact that the spectral function of the intermediate unstable particle does not have an isolated pole but a continuum part, Ref.~\cite{Ala-Mattinen:2023rbm} proposes to rescale its classical phase space density $f$ via $f \to R(\delta)f$ with the one-particle weight function
\begin{equation}
    R(\delta) = \frac{1}{\pi} \int \dd{s}\, \Theta(s-m^2,\delta)\, m\Gamma \, |\Delta_{\rm BW}(s)|^2,
\end{equation}
that is constrained to be less than unity. These definitions are made to 1) avoid by definition unphysical negative cross sections that might appear in the SDF and PVS schemes~\cite{Cline:2017qpe} through an overestimation of the subtracted on-shell part, 2) propose a scheme that is also applicable to interference matrix elements, and 3) allow one to estimate through variation of $\delta$ for which decay widths the particle picture of the intermediate state still holds. While $\delta = m \Gamma$ might seem like a natural choice, this scheme introduces an artificial dependence on $\delta$.

\item \textit{\texttt{DarkSUSY}-scheme:} 
Starting from the requirement that the squared Breit--Wigner propagator reproduces within the narrow-width approximation close to the resonance the $1\to 2$ collision term describing the mediator decay into DM, the DM tool \texttt{DarkSUSY}~\cite{Bringmann:2018lay, Bringmann_2022} recommends --- at least for a single scalar mediator --- the manual modification of the partial vacuum decay width $\Gamma^{\rm vac}_{Y\to A,B}$ inside the Breit--Wigner propagator according to 
\begin{equation}
    \Gamma_{\rm eff}^{\texttt{DarkSUSY}} = \frac{1}{1 + f_Y(\omega)}\sum_{A,B} \Gamma^{\rm vac}_{Y\to A,B} \, \bar{G}_{AB}(\mathbf{p}, m_Y^2) \, ,
    \label{eq:GammaDarkSUSY}
\end{equation}
where $\omega=\sqrt{\mathbf{p}^2 + m_Y^2}$ is the mediator's energy in the cosmic rest frame and the sum runs over all particles $A$, $B$ from the thermal bath. The suppression factor $1/(1 +f_Y)$ was added with the finite-temperature QFT cutting rules~\cite{PhysRevD.28.2007-thermalcuttingrules} in mind to account for the fact that the imaginary part of the mediator self-energy is given by the difference of the decay and inverse decay rates. Eq.~\eqref{eq:GammaDarkSUSY} holds strictly only for two-body decays. However, NLO vacuum widths as used in Ref.~\cite{Bringmann_2022} also involve real emissions, which lead to additional phase-space densities not captured by $\bar{G}_{AB}$.

\item \textit{\texttt{micrOMEGAs}-scheme:} 
The \texttt{micrOMEGAs} package~\cite{Belanger:2018ccd}\footnote{This definition of the \texttt{micrOMEGAs}-scheme is valid at least up to the most recent \texttt{micrOMEGAs} version \texttt{7.1.4} available at the time of writing this paper.} for the computation of DM observables in generic models treats the overcounting issue by discarding the decay collision term through a modification of the Breit--Wigner width in the denominator as well. However, in contrast to Eq.~\eqref{eq:GammaDarkSUSY}, the function $\bar{G}_{AB}(\mathbf{p}, m_Y^2)$ is not evaluated for the three-momentum going into the propagator, but averaged according to the momentum distribution of the mediator in a first step to yield an average partial decay width of the mediator into bath particles,
\begin{equation}
     \Gamma^{\mathrm{bath}}_{\mathrm{tot}} = \sum_{A,B}\frac{\int \frac{\dd[3]{\mathbf{p}}}{2\omega} \, f_Y(\omega) \, \bar{G}_{AB}(\mathbf{p}, m_Y^2)}{\int \frac{\dd[3]{\mathbf{p}}}{2\omega} \, f_Y(\omega)}\, \Gamma^{\rm vac}_{Y\to A,B} \, .
     \label{eq:MO_bath}
\end{equation}
Then, it is taken into account that, for a small mediator width, $Y$ decays not instantaneously, but rather at a ``decay'' temperature $T_d$ smaller than its production temperature according to the survival probability $P(T_d) = e^{-L(T_d)}$ with the exponent
\begin{equation}
   L(T_d) =  \int_{T_d}^T \dd{T'}\, \frac{  \Gamma^{\mathrm{bath}}_{\mathrm{tot}}(T')}{\gamma(T')\overline{H}(T')T'}\,.
\end{equation}
Here, the modified Hubble rate is defined as 
\begin{align}
    \overline{H}(T) = \frac{H(T)}{1+\frac{1}{3}\dv{ \!\ln{h_\mathrm{eff}}}{\!\ln{T}}},
\end{align}
with $h_\text{eff}(T)$ denoting the effective number of entropy degrees of freedom, while the purely conventional boost factor is given by
\begin{align}
    \gamma(T) = \frac{\int \dd[3]{\mathbf{p}}\, f_Y(\omega,T)}{\int \dd[3]{\mathbf{p}}\frac{m_Y}{\omega} \, f_Y(\omega,T)} \, .
\label{eq:avboost}
\end{align}
Averaging over the inverse bath decay rate Eq.~\eqref{eq:MO_bath} using $P(T_d)$ from production at the bath temperature $T$ until today's photon temperature $T_0$ yields the effective width
\begin{equation}
    \Gamma_{\rm eff}^{\texttt{micrOMEGAs}} =   \frac{\int \frac{\dd[3]{\mathbf{p}}}{2\omega} \, e^{-\frac{\omega}{T}} \, \bar{G}_{AB}(\mathbf{p}, m_Y^2)}{\int \frac{\dd[3]{\mathbf{p}}}{2\omega} \, f_Y(\omega) \, \bar{G}_{AB}(\mathbf{p}, m_Y^2)} \bigg/  \left(\int_{T_0}^{T} \dd{T_d} \, \frac{P(T_d)}{\gamma(T_d)\overline{H}(T_d)T_d}\right)
    \label{eq:MO_eff}
\end{equation}
used for the mediator width in the $A,B\to Y^\ast \to \text{DM}, \overline{\text{DM}}$ annihilation process. The prefactor, a ratio of Maxwell–Boltzmann and quantum-statistical averages of $\bar{G}_{AB}$, accounts for a potential mediator chemical potential, so that Eq.~\eqref{eq:MO_eff} must, unlike Eq.~\eqref{eq:GammaDarkSUSY}, be recomputed for each annihilation channel. The width Eq.~\eqref{eq:MO_eff} is then used with \texttt{CalcHEP}'s default treatment~\cite{Pukhov:1999gg,Pukhov:2004ca,Belyaev:2012qa} of Breit--Wigner propagators and their interference terms with other resonant and non-resonant propagators designed to keep the gauge-motivated cancellation of large terms between different diagrams.\footnote{For the example models and parameter regions of Sec.~\ref{sec: Numerical Examples}, we have verified numerically that solving the DM evolution equation with the effective width Eq.~\eqref{eq:MO_eff} in our own code reproduces the \texttt{micrOMEGAs} relic densities within $\SI{0.1}{\percent}$.}

\end{itemize}

\section{Freeze-In in the CTP Formalism}
\label{sec:Freeze-in in the CTP Formalism}

To compute observables in non-equilibrium systems from first principles in Quantum Field Theory, the former are typically translated into expectation values of operators that are not time ordered with respect to an at least initially known statistical density matrix. While in equilibrium such expectation values may be obtained using the Matsubara imaginary-time formalism~\cite{Matsubara:1955ws,LeBellac_1996,Laine:2016hma}, the Keldysh--Schwinger or CTP formalism~\cite{10.1063/1.1703727-julianschwinger,Keldysh:1964ud,Danielewicz:1982kk,Chou:1984es,Calzetta:1986cq,Rammer:1986zz,Botermans:1990qi,Berges_2004} extends as a real-time framework also to systems out of equilibrium and is therefore well-suited for DM freeze-in production, where the dark sector starts from zero initial abundance. The main expectation values of interest for us are two-point functions that we generically label as $G^{ab}$ with $a,b\in\{1,2\}$ being CTP contour indices, since these encode different thermodynamic properties of the system under consideration, in particular the DM number density. To discriminate different field types we introduce $\Delta^{ab}$ as the propagator for scalar fields and analogously $\slashed S^{ab}$ and $D^{ab}_{\mu\nu}$ for fermionic and vector fields, respectively. Their corresponding free forms in equilibrium as well as their relations among each other are listed as a quick reference in App.~\ref{appendix: ctp props}.

Starting from the CTP formalism, we derive in the following from the afore-mentioned two-point functions for all three field types local kinetic equations for the evolution of the DM number density applicable to freeze-in production that do not overcount mediator decays. 

\subsection{Kadanoff--Baym Equations and Approximations}

The equations of motion for the non-equilibrium two-point functions on the complex time contour are the Dyson--Schwinger equations. For a scalar field $S$ with position-space self-energy $\Pi_S(x,y)$ these read, for example,
\begin{align}
     (\Box_x + m_S^2) \Delta_S(x,y) = -\int_C \dd[4]{z} \Pi_S(x,z) \Delta_S(z,y) - \delta_C^{(4)}(x,y) 
     \label{eq:DS}
\end{align}
with the complex time contour $C$ running from $-\infty$ to $\infty$ and back.
By subdividing them into their four real-time analogues one obtains the Kadanoff--Baym equations \cite{Baym:1961zz,KadanoffBaym:1962}. For a scalar field $S$, a {Majorana or Dirac} fermion $\chi$, and a massive vector field $V_\mu$ described by a Proca Lagrangian, i.e., working in unitary gauge, and denoting the corresponding fermionic and vector CTP self-energies by $\Sigma_\chi^{ab}$ and $\Pi^{ab}_{V,\mu\nu}$, respectively, the Kadanoff--Baym equations for the corresponding Wightman two-point functions read
\begin{subequations}\label{eq:KB_equations}
\begin{align}
\left(\partial^2 + m_S^2\right)\Delta_S^{\lessgtr}+\Pi_S^{\mathcal H}\odot \Delta_S^{\lessgtr}+\Pi_S^{\lessgtr}\odot \Delta_S^{\mathcal H}&=\frac{1}{2}\left(\Pi_S^{<} \odot \Delta_S^{>}-\Pi_S^{>} \odot \Delta_S^{<}\right),
\\
\left(i\slashed{\partial}-m_\chi\right)\slashed S_\chi^{\lessgtr}-\slashed \Sigma_\chi^{\mathcal H}\odot \slashed S_\chi^{\lessgtr}-\slashed \Sigma_\chi^{\lessgtr}\odot \slashed S_\chi^{\mathcal H}
&=
\frac{1}{2}\left(\slashed \Sigma_\chi^{>} \odot \slashed S_\chi^{<}-\slashed \Sigma_\chi^{<} \odot \slashed S_\chi^{>}\right),
\\
L_{\mu\rho}D_V^{\rho\nu,\lessgtr}-\Pi_{V,\mu\rho}^{\mathcal H}\odot D_V^{\rho\nu,\lessgtr}-\Pi_{V,\mu\rho}^{\lessgtr}\odot D_V^{\rho\nu,\mathcal H}&=\frac{1}{2}\left(\Pi_{V,\mu\rho}^{>} \odot D_V^{\rho\nu,<}-\Pi_{V,\mu\rho}^{<} \odot D_V^{\rho\nu,>}\right).
\label{eq:FKB_vector}
\end{align}
\end{subequations}
All differential operators in Eq.~\eqref{eq:KB_equations} act implicitly on the first space-time argument. The kinetic operator for the Proca field is $L_{\mu{\nu}}=
\eta_{\mu{\nu}}\left(-\partial^2 - m_V^2\right)+\partial_\mu\partial_{\nu}$ with the Minkowski metric $\eta_{\mu\nu}=\operatorname{diag}(1,-1,-1,-1)$ and the convolution $(A\odot B)(x,y)=\int \mathrm d^4z\,A(x,z)B(z,y)$. Since the full Kadanoff--Baym equations are non-local integro-differential equations, a general analytic closed-form solution is currently not known. To derive local kinetic equations, we introduce the mean and relative coordinates $X=\frac12(x+y)$ and $r=x-y$, respectively, and perform a Wigner transformation,
\begin{align}
    G(X,p)=\int \dd[4]{r} \, e^{ip\cdot r} G\left(X+\frac{r}{2}, X-\frac{r}{2}\right).
\end{align}
Under this transformation, the convolutions $\odot$ in the Kadanoff--Baym equations are mapped onto Moyal products,
\begin{align}
    (A \odot B)(X,p)= A(X,p) \exp\left[-\frac i 2 \left(\overleftarrow{\partial}_X \cdot \overrightarrow{\partial}_p - \overleftarrow{\partial}_p \cdot \overrightarrow{\partial}_X\right)\right]B(X,p),
\end{align}
where the left- and right-pointing arrows indicate derivatives acting on $A$ and $B$, respectively. The Moyal products can be expanded in gradients with respect to the slowly varying macroscopic coordinate $X$~\cite{Danielewicz:1982kk,Rammer:1986zz,Calzetta:1986cq,Botermans:1990qi}.
To make the range of validity of this procedure explicit, we use the following assumptions:

\begin{enumerate}
    \item \textit{Homogeneity and isotropy.} We work in the cosmic rest frame and assume the plasma to be {spatially} homogeneous and isotropic. This allows us to neglect spatial gradients with respect to the mean coordinate $X$, such that the two-point functions depend on $X$ only through the macroscopic time variable $t=X^0$.
\item \textit{Equilibrium bath.}
The thermal bath, including the mediator, is assumed to remain in equilibrium during DM production. The corresponding fields are therefore described by equilibrium CTP two-point functions that satisfy the Kubo--Martin--Schwinger (KMS) relation. In contrast, the DM field is not assumed to thermalize and is treated as a non-equilibrium degree of freedom.
\item \textit{Kadanoff--Baym and quasi-particle ansatz for DM.} We adopt the Kadanoff--Baym ansatz, i.e., a generalized fluctuation--dissipation relation~\cite{Garny:2009rv,Garbrecht:2011xw,BhupalDev:2014oar,Drewes_2013-KadanoffBaymAnsatz} that decomposes the DM Wightman functions
\begin{subequations}
\begin{align}
    i\Delta^<_S({t},p) &= 2\Delta_S^\mathcal A(p)\,f_S({t},\mathbf p)\, , & i\slashed S_\chi^<(t,p) &= -2\slashed S_\chi^\mathcal A(p)\,f_\chi({t},\mathbf p)\, ,
    \\
    i\Delta_S^>({t},p) &= 2\Delta_S^\mathcal A(p)\left[1+f_S({t},\mathbf p)\right]\, , & i\slashed S_\chi^>(t, p) &= 2\slashed S_\chi^\mathcal A(p)\left[1-f_\chi({t},\mathbf p)\right]\, ,
\end{align}
    \label{eq:KBansatz}
\end{subequations}
into the anti-Hermitian two-point functions $G^\mathcal A$ and the corresponding DM distribution functions $f_{\rm DM}$.
Since the DM couplings are feeble, we follow the quasi-particle ansatz and approximate their spectral functions with their zero-width forms, 
\begin{subequations}
\begin{align}
    \Delta_S^\mathcal A(p) &= \pi\,\operatorname{sgn}(p_0)\, \delta(p^2-m_S^2),
    \\
    \slashed S_\chi^\mathcal A(p) &= \pi\,\operatorname{sgn}(p_0)\, (\slashed p+m_\chi)\, \delta(p^2-m_\chi^2).
\end{align}
\label{eq:QPA}
\end{subequations}
For the vector DM Wightman function our ansatz proceeds analogously. 

\item \textit{Freeze-in limit.} The DM phase-space density {is assumed to be} much smaller than unity, \(f_{\rm DM}\ll1\). We therefore neglect {all $\order{f_{\rm DM}}$ and higher-order contributions associated with any internal DM line within the perturbative weak-coupling expansion of the DM self-energy. 
This also corresponds to the neglect of inverse processes with DM in the initial state. 
This also justifies the usage of standard finite-temperature perturbation theory~\cite{Danielewicz:1982kk, Landsman:1986uw, Niemi:1983nf, Ali:2026ehk} despite the DM candidate being out-of-equilibrium.} 

\item \textit{Leading-gradient approximation.} 
We truncate the gradient expansion of the Moyal products at leading order and discard contributions that are schematically of the form $\mathcal O\left(\partial_t\Pi^{\mathcal H,\lessgtr}\Delta^{\mathcal H,\lessgtr}\right)$.
\end{enumerate}
With these assumptions, the Kadanoff--Baym equations reduce to a set of local kinetic equations 
\begin{subequations}\label{eq: kinetic equations}
\begin{align}
    p_0 \partial_t (i\Delta_S^{\lessgtr}) &=-\frac{1}{2}\left(i\Pi_S^{>} i\Delta_S^{<}-i\Pi_S^{<} i\Delta_S^{>}\right)+\mathcal{O}\!\left(\partial_t \Pi_S^{\mathcal{H},\lessgtr}\Delta_S^{\mathcal{H},\lessgtr}\right),
\label{eq:prop_kinetic_equation_scalar_truncated}
\\
\gamma^0 \partial_t \bigl(-i\slashed{S}_\chi^{\lessgtr}\bigr)&=i\slashed{\Sigma}_\chi^{>} \, i\slashed{S}_\chi^{<}-i\slashed{\Sigma}_\chi^{<} \, i\slashed{S}_\chi^{>}+\mathcal{O}\!\left(\partial_t \slashed{\Sigma}_\chi^{\mathcal{H},\lessgtr} \,\slashed{S}_\chi^{\mathcal{H},\lessgtr}\right),
\label{eq:prop_kinetic_equation_fermion_truncated}
\\
p_0 \partial_t (iD_V^{\mu\nu,\lessgtr}) &= -\frac{1}{2} \left(i \Pi_V^{>\,\mu}{}_{\rho}\,iD_V^{\rho\nu,<} - i\Pi_V^{<\,\mu}{}_{\rho}\,iD_V^{\rho\nu,>} \right) + \mathcal{O}\!\left[ \left( \partial_t\Pi_V^{\mathcal H,\lessgtr} D_V^{\mathcal H,\lessgtr} \right)^{\mu\nu} \right]
\label{eq:prop_kinetic_equation_vector_truncated}
\end{align}
\end{subequations}
for the Wightman functions, that we turn into evolution equations for the DM number density in the next subsection. 

\subsection{Evolution Equation for the Dark Matter Number Density} \label{sec: Evolution Equation for the Dark Matter Density}

Starting with scalar DM $S$, we now turn the kinetic Eqs.~\eqref{eq: kinetic equations} into evolution equations for the DM number density. The straightforward extension to fermionic $\chi$ and vector DM $V$ follows below. Using the Kadanoff--Baym Eq.~\eqref{eq:KBansatz} and the quasi-particle ansatz Eq.~\eqref{eq:QPA} for the DM two-point function and inserting it into the kinetic equation, one obtains an evolution equation for the scalar DM distribution function,
\begin{align}
\partial_t f_S(t,\bfp)\, &\overset{\mathclap{\eqref{eq:KBansatz}}}{=}\, \int_0^\infty \frac{\dd{p^0}}{\pi}\, p_0\,\partial_t i\Delta_S^<(p) \nonumber\\ &\overset{\mathclap{\eqref{eq:prop_kinetic_equation_scalar_truncated}}}{=}\, \int_0^\infty \frac{\dd{p^0}}{\pi}\, \frac{1}{2} \left( i\Pi_S^<(p)i\Delta_S^>(p) - i\Pi_S^>(p)i\Delta_S^<(p) \right)
\nonumber\\
&\overset{\mathclap{\eqref{eq:KBansatz},\eqref{eq:QPA}}}{=}\quad \int_0^\infty\dd{p^0}\, \operatorname{sgn}(p^0)\, \delta(p^2-m_S^2) \left[ i\Pi_S^<(p)(1+f_S) - i\Pi_S^>(p)f_S \right] \nonumber\\ &\simeq\, \frac{i\Pi_S^<(\omega_p,|\mathbf p|)}{2\omega_p}, 
\label{eq:scalar_dm_distribution_evolution}
\end{align}
where \(\omega_p=\sqrt{|\mathbf p|^2+m_S^2}\) is the DM energy. The last line follows from the positive-energy on-shell contribution and from the freeze-in condition $f_S\ll1$, which allows us to neglect terms proportional to the DM phase-space density. Note that Eq.~\eqref{eq:scalar_dm_distribution_evolution} is expressed through the Wightman function $i\Pi_S^<$ rather than the spectral self-energy. If all fields running in the loop of the DM self-energy are in equilibrium, the KMS relation $i\Pi_S^< = 2 f_\text{BE}(\omega_p)\,\Pi_S^{\mathcal{A}}$ allows one to rewrite Eq.~\eqref{eq:scalar_dm_distribution_evolution} in terms of $\Pi_S^{\mathcal{A}}$, as done in Ref.~\cite{Becker:2023vwd}. For $s$-channel pair production, however, one internal line of the self-energy in Fig.~\ref{fig:oneloopresummed} is itself a DM propagator, which is out of equilibrium. The KMS relation then does not hold for $\Pi_S$, and the lesser ($<$) self-energy has to be used directly.
Embedding this local Minkowski-space rate, evaluated in the plasma rest frame, into an FLRW background adds the usual Hubble dilution $3Hn_S$.

Applying the usual definition of the number density for symmetric DM\footnote{Symmetric is meant in contrast to the idea of asymmetric DM~\cite{Petraki:2013wwa}.}
\begin{align}
    n_{\rm DM}(t)=\kappa\, g_{\rm DM} \int \frac{\dd[3]{\bfp}}{(2\pi)^3}\, f_{\rm DM}(t,\bfp)
\end{align}
with $\kappa=2(1)$ for non-self-conjugate (self-conjugate) DM and where $g_{\rm DM}=1,2,3$ denotes the number of spin states for scalar, spin-$1/2$ and massive vector DM, respectively,  gives, after the integration of  Eq.~\eqref{eq:scalar_dm_distribution_evolution} over the physical three-momentum,
\begin{align}\label{eq: scalar DM evolution equation}
\partial_t n_S(t)+3Hn_S(t) &= \gamma_{S} \equiv \frac{ \kappa}{4\pi^2} \int_{m_S}^\infty \dd{\omega_p} |\mathbf p| \, i\Pi_S^<(\omega_p,|\mathbf p|).
\end{align}
For fermionic and vector DM, the respective phase-space distributions are obtained from the Wightman functions as
\begin{equation}
    f_\chi(t,\bfp) = \int_0^\infty \frac{\dd{p^0}}{4\pi} \Tr\left[-i\slashed S_\chi^<(t,p)\gamma^0\right],\qquad f_V(t,\bfp)= \int_0^\infty \frac{\dd p^0}{3\pi} p^0\, P_{\mu\nu}(p)\,iD^{\mu\nu, <}_V(t,p),
\end{equation}
with the polarization sum $P_{\mu\nu}(p)=-\eta_{\mu\nu}+\frac{p_\mu p_\nu}{m_V^2}$.
Under the same assumptions as in the scalar case, the number density evolution equations turn out to be
\begin{subequations} \label{eqs: fermionic and vector dm}
\begin{align}
\partial_t n_\chi(t)+3Hn_\chi(t)=& \gamma_\chi \equiv \frac{ \kappa }{4\pi^2}\int_{m_\chi}^\infty \dd{\omega_p}|\mathbf p| \eval{\operatorname{Tr}[-i\slashed{\Sigma}_\chi^<(p)(\slashed p  +m_\chi)]}_{p_0=\omega_{p}},
\\
\partial_t n_V(t)+3Hn_V(t) =& \gamma_V \equiv \frac{ \kappa}{4\pi^2} \int_{m_V}^\infty \dd{\omega_p} |\mathbf p|   \eval{\left[-iP_{\mu\nu}(p)\Pi_V^{<~\mu\nu}(p)\right]}_{p_0=\omega_{p}}\, .
\end{align}
\end{subequations}
In all three cases, the right-hand side of the number-density equation can be interpreted as the CTP collision term that we denote as the DM production rate $\gamma_{\rm DM}$.
It is obtained by integrating the lesser DM self-energy projected onto the corresponding physical on-shell states.

\section{1PI-Resummed CTP Treatment of $\mathbf{s}$-Channel Freeze-In}
\label{sec:CTP-Resummed}

We now apply this result to $s$-channel pair production of DM via a scalar or vector mediator in equilibrium and express the production rate through the finite-width mediator spectral density that combines the resonant and off-shell contributions in a single expression. The analysis is subsequently specialized to a one-loop mediator self-energy generated by two equilibrium particles $A$ and $B$. In the kinematic regime where this contribution corresponds to mediator production through bath-particle annihilation, the general spectral-density description can be recast in the form of a Boltzmann-like $2\to2$ collision term for the reaction $A,B\to Y^\ast \to \text{DM}, \overline{\text{DM}}$.
    
\subsection{CTP Production Rate}

    \begin{figure}[t]
        \centering
        \includegraphics[width=0.5\linewidth]{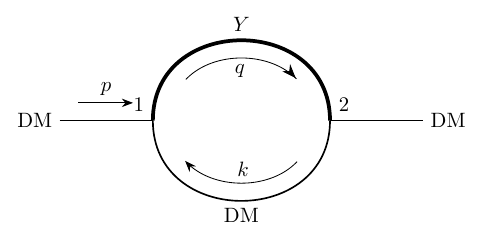}
        \caption{DM self-energy topology associated with the $s$-channel CTP production rate. The internal DM line is evaluated in the quasi-particle approximation, while the thick line denotes the 1PI-resummed mediator propagator.}
        \label{fig:oneloopresummed}
    \end{figure}
    
We consider a generic $s$-channel freeze-in process in which DM pair production proceeds via the CTP diagram shown in Fig.~\ref{fig:oneloopresummed} and neglect all higher-order contributions in the DM-mediator coupling.
For all three DM types, the associated CTP production rate, derived in Sec.~\ref{sec: Evolution Equation for the Dark Matter Density}, can be written in the generic form
\begin{align}
\gamma_\text{DM} = \frac{2}{S_{\rm DM}} \int \dd \Pi_p\,\dd \Pi_k\, \mathcal H_{\rm DM}^{IJ}(p,k)\, iG^<_{Y,IJ}(p+k), \label{eq:ctp-collision-general} 
\end{align}
where $S_{\rm DM}$ denotes the symmetry factor of the two-particle dark-sector final state with $S_{\rm DM}=2$ for identical particles and $S_{\rm DM}=1$ otherwise. The explicit factor of two accounts for the two DM particles produced per pair-production event, such that $\gamma_\text{DM}$ gives the CTP analogue of the Boltzmann collision term Eq.~\eqref{Eq: BoltzmannEq}. The tensor $\mathcal H_{\rm DM}^{IJ}$ is the final-state tensor associated with the decay $Y^\ast \to \text{DM}, \overline{\text{DM}}$. It contains the squared coupling associated with the DM-mediator interaction, together with the DM spin sums. For a scalar mediator, the indices $I,J$ are absent, whereas for a vector mediator they correspond to Lorentz indices. For example, for fermionic DM coupled to a vector mediator $Z'$ through $\mathcal L_{\rm int} \supset g_\chi Z'_\mu\bar\chi\gamma^\mu\chi$,  one obtains
\begin{align} 
\mathcal H_\chi^{\mu\nu} = g_\chi^2 \operatorname{Tr} \left[ (\slashed p+m_\chi)\gamma^\mu (\slashed k-m_\chi)\gamma^\nu \right]. 
\end{align}
In Eq.~\eqref{eq:ctp-collision-general}, the properties of the mediator are encoded in its lesser Wightman function. For a mediator in thermal equilibrium, the KMS relation separates the statistical occupation from the spectral information, $iG_{Y,IJ}^<(q)=f_Y(q_0)\rho_{Y,IJ}(q)$.

\subsection{Mediator Spectral Densities}

The spectral density of the resummed mediator is determined by the Hermitian and anti-Hermitian components of its self-energy, $\Pi_Y^{\mathcal H}$ and $\Pi_Y^{\mathcal A}$, respectively. The Hermitian part modifies the dispersive properties of the mediator, while the anti-Hermitian component describes the absorptive part and determines its in-medium width. The explicit form of the spectral density depends on the spin of the mediator. We therefore discuss scalar and vector mediators separately.

\paragraph{Scalar mediator}
For a scalar mediator, the resummed propagators derived in App.~\ref{sec:app:ctp-resummed-propagator} yield the spectral density
\begin{align}
\rho_Y(q) = \frac{2\Pi_Y^{\mathcal A}(q)}{\left[q^2-m_Y^2-\Pi_Y^{\mathcal H}(q)\right]^2 + \left[\Pi_Y^{\mathcal A}(q)\right]^2}\, ,
\label{eq:scalar-spectral-density}
\end{align}
where $m_Y$ is the physical vacuum mass and we consider only the temperature-dependent part of the Hermitian self-energy. Let us note that Eq.~\eqref{eq:scalar-spectral-density} would even be valid for an out-of-equilibrium scalar mediator if the backreaction of the field $Y$ is neglected~\cite{Anisimov:2008dz}. In the vacuum limit, $\Pi_Y^{\mathcal A}$ reduces to the absorptive vacuum self-energy associated with the usual decay width. 
For a scalar mediator, the Lorentz indices in Eq.~\eqref{eq:ctp-collision-general} are absent. After performing the two-body phase-space integral over the DM final state at fixed mediator momentum $q$, with $s=q^2$, and expressing the result in terms of the off-shell partial vacuum decay-width $\Gamma^{\rm vac}_{Y^\ast \to \text{DM},\overline{\text{DM}}}(s)$, the production rate becomes

\begin{empheq}[box=\eqbox]{align}\label{eq: general collsionterm CTP scalar}
    \gamma_\text{DM}=4\int \frac{\dd[4]{q}}{(2\pi)^4}  \sqrt{s}\,\Gamma^{\rm vac}_{Y^\ast \to \text{DM},\overline{\text{DM}}}(s)f_Y(q_0)\rho_Y(q).
\end{empheq}

\paragraph{Vector mediator}
At finite temperature and in an isotropic medium, the CTP vector mediator self-energy $\Pi^{ab}_{\mu\nu}$ is no longer a scalar as before but a rank-2 tensor constructed out of the Minkowski metric $\eta^{\mu\nu}$, the virtual-mediator momentum $q^\mu$, and the four-velocity of the heat bath $u^\mu$. Therefore, $\Pi^{\mu\nu}$ is in general a linear combination of five tensors. One common choice is to take the thermal bath at rest, i.e., $u_\mu=(1,\mathbf 0)$, and use the basis~\cite{Nieves:1988qz,Buchmuller:1993bq,Tang:2019kwx}
\begin{subequations}\label{Eq: projectorsVectorMediator} 
\begin{align}
&P_\mathrm T^{00}=P_\mathrm T^{0i}=P_\mathrm T^{i0}=0, \quad P_\mathrm T^{ij}=\delta^{ij}-\frac{\mathbf q^i\mathbf q^j}{|\mathbf q|^2},\\
&P_\mathrm L^{\mu\nu}=-\eta^{\mu\nu}+\frac{q^\mu q^\nu}{q^2}-P_\mathrm T^{\mu\nu}\, , \quad P_ \mathrm G^{\mu\nu} = \frac{q^\mu q^\nu}{q^2}\, , \quad P^{\mu\nu}_\mathrm A = \epsilon^{\mu\nu\alpha\beta} q_\alpha u_\beta \, ,\\ &S^{\mu\nu} = \frac{1}{\sqrt{2}|\mathbf{q}|} \left(q^\mu \tilde u^\nu + q^\nu \tilde u^\mu\right)\, ,
\end{align}
\end{subequations}
where the auxiliary direction $\tilde u^\mu = u^\mu - \frac{u\cdot q}{q^2}q^\mu$ is transverse, $\tilde u \cdot q =0$. {For clarity, CTP indices are left implicit in the decomposition of the mediator self-energy,}
\begin{equation}
     \Pi^{\mu\nu} = P^{\mu\nu}_\mathrm T \Pi_\mathrm T + P^{\mu\nu}_\mathrm L \Pi_\mathrm L + P^{\mu\nu}_\mathrm G \Pi_\mathrm G +P^{\mu\nu}_\mathrm A \Pi_\mathrm A + S^{\mu\nu} \Pi_\mathrm S .
\end{equation}
The anti-symmetric term $\Pi_\mathrm A$ receives only support if there are $P$ or $CP$-odd effects in the Lagrangian or the background medium~\cite{Nieves:1988qz}. For simplicity, we therefore consider in the following only $P$ and $CP$-symmetric theories where the mediator self-energy is in addition solely generated by a conserved current such that the Ward identity $q_\mu \Pi^{\mu\nu}=0$ holds and consequently only the transverse and longitudinal components, $\Pi_\mathrm T$ and $\Pi_\mathrm L$, are non-vanishing. This is because $P_\mathrm T^{\mu\nu}$ describes the two polarizations transverse to $\mathbf q$ that are also present in vacuum, while $P_\mathrm L^{\mu\nu}$ accounts for longitudinal excitations generated through interactions with the medium or a mass term in the Lagrangian. The transversality requirement naturally includes unbroken Abelian~\cite{Weldon:1996kb} gauge theories, while one example theory with a massive vector mediator compatible with this condition will be given in Sec.~\ref{sec:vetor-model}. Due to the orthogonality of the transverse and longitudinal projection tensors and their closure under contraction, the two polarization sectors decouple and can be resummed independently. Analogously to the scalar mediator case, both spectral densities take the form
\begin{align} \label{eq: spectral density vectormediator}
\rho_{Y,i}(q) = \frac{ 2\Pi_{Y,i}^{\mathcal A}(q) }{ \left[ q^2-m_Y^2-\Pi_{Y,i}^{\mathcal H}(q) \right]^2 + \left[ \Pi_{Y,i}^{\mathcal A}(q) \right]^2 }, \qquad i=\mathrm T,\mathrm L. 
\end{align}
Performing again the two-body phase-space integral over the DM final state at fixed $q$ and expressing the result in terms of the off-shell decay width $\Gamma^{\rm vac}_{Y^\ast \to \text{DM},\overline{\text{DM}}}(s)$, the production rate becomes
\begin{empheq}[box=\eqbox]{align}
\label{eq_ general collsion term vectormediator}
    \gamma_{\rm DM}=4 \int \frac{\dd[4]{q}}{(2\pi)^4}   \, \sqrt{s}\, \Gamma^{\rm vac}_{Y^\ast \to \text{DM},\overline{\text{DM}}}(s)\, f_Y(q_0)\, \left[2  \rho_{Y,\mathrm T}(q) + \rho_{Y,\mathrm L}(q)\right].
\end{empheq}
The production rates Eqs.~\eqref{eq: general collsionterm CTP scalar} and \eqref{eq_ general collsion term vectormediator} constitute one of the main results of this work. Together with the evolution equation 
\begin{empheq}[box=\eqbox]{align}
   \dot{n}_{\rm DM} + 3 H n_{\rm DM} = \gamma_{\rm DM}
\end{empheq}
they allow to self-consistently describe freeze-in production of scalar, fermion and vector DM via a bosonic mediator.

\subsection{One-Loop {Absorptive} Mediator Self-energy}
\label{sec: One-loop bath contribution}

We now derive generic expressions for the one-loop anti-Hermitian mediator self-energy $\Pi_Y^{\cal A}(q)$ from its coupling to two bath particles $A$ and $B$ in the kinematic region where $A$ and $B$ are produced simultaneously on shell. This requires $q^0>0$ and $s=q^2\geq (m_A+m_B)^2$. The Hermitian self-energy on the other hand has to be evaluated on a case-by-case basis with example calculations in App.~\ref{sec: app: hermitian self-energys}.

The anti-Hermitian self-energy may be extracted from the greater one-loop self-energy together with the KMS relation $i\Pi_Y^>(q)=2\bigl[1+f_Y(q_0)\bigr]\Pi_Y^{\mathcal A}(q).$ For the bath particles $A$ and $B$, this gives
\begin{multline}\label{eq: general antihermiitian selfenergy}
\Pi_{Y}^{\mathcal A\,IJ}(q) = \frac{(2S_{AB})^{-1}}{1+f_Y(q_0)} \int\dd{\Pi_A}\dd{\Pi_B}\,(2\pi)^4\delta^{(4)}(q-p_A-p_B)
\\
\times [1 + \eta_A f_A(E_A)][1 + \eta_B f_B(E_B)]\mathcal H_{\rm bath}^{IJ},
\end{multline}
where $S_{AB}$ is the symmetry factor of the bath particles and $\mathcal H_{\rm bath}^{IJ}$ contains the coupling factors associated with the $YAB$ interaction together with the spin or polarization sums of the bath particles $A$ and $B$. 

\paragraph{Scalar mediator}
For a scalar mediator, the two-body phase space can be directly related to the corresponding vacuum two-body decay width of a mediator with invariant mass $\sqrt{s}$,
\begin{align}
\Gamma^{\mathrm{vac}}_{Y^\ast \to A, B}(s)=\frac{1}{2\sqrt{s}}\frac{1}{S_{AB}}\int\mathrm{d}\Pi_A\mathrm{d}\Pi_B(2\pi)^4\delta^{(4)}(q-p_A-p_B)\mathcal{H}_{\mathrm{bath}}(p_A,p_B).
\label{eq:vacuum-width}
\end{align}
Using the thermal phase-space factor $\bar G_{AB}(\mathbf q,s)$ introduced in Eq.~\eqref{eq:Gab}, the anti-Hermitian self-energy can be written as
\begin{align}\label{eq:scalar-thermal-PiA}
\Pi_Y^{\mathcal A}(q) = \frac{ \sqrt{s}\,\Gamma^{\mathrm{vac}}_{Y^\ast \to A,B}(s)}{1+f_Y(q_0)} \bar{G}_{AB}(\mathbf{q},s).
\end{align}
An alternative analytic solution is, e.g., available in Ref.~\cite{Drewes:2013bfa}. For the scalar mediator, the thermal modification of the anti-Hermitian self-energy is therefore entirely encoded in the factor $\bar G_{AB}/[1+f_Y]$. That is the same factor that appears in the \texttt{DarkSUSY} prescription of Eq.~\eqref{eq:GammaDarkSUSY}. In the vacuum limit, $\bar G_{AB}\to 1$ and $f_Y\to 0$, such that $\Pi_Y^\mathcal A(s,T=0)=\sqrt s\,\Gamma ^{\rm vac}_{Y^\ast \to A,B}(s).$

\paragraph{Vector mediator}
For a vector mediator, its anti-Hermitian self-energy has to be projected onto the transverse and longitudinal modes using the projectors introduced in Eq.~\eqref{Eq: projectorsVectorMediator}. In contrast to the scalar case, the contraction of the bath tensor with the corresponding projection operators introduces an angular dependence on $\cos\theta^{\rm CM}$. The thermal phase-space integral consequently does no longer factorize into the function $\bar G_{AB}$ and the decay rate $\Gamma_{Y^\ast\to A,B}^{\rm vac}(s)$. Therefore, we instead define the normalized weight functions 
\begin{align}
W_i(\cos\theta^{\rm CM})=\frac{ P_i^{\mu\nu}(q)\mathcal H_{\rm{bath},\mu \nu}(p_A(\cos \theta^{\rm CM}),p_B(\cos \theta^{\rm CM}))}{\int_{-1}^1\frac{\dd\cos\theta^{\rm CM}}{2}P_i^{\mu\nu}(q)\mathcal H_{\rm{bath},\mu \nu}(p_A(\cos \theta^{\rm CM}),p_B(\cos \theta^{\rm CM})) }\, , \quad i=\mathrm T,\mathrm L \, ,
\end{align}
motivated by the fact that, in the zero-temperature limit, as a result of the restored Lorentz invariance, the longitudinal and transverse components are no longer independent scalar functions, but yield identically the vacuum width, $\Pi_{Y,\mathrm T/\mathrm L}^\mathcal A(s,T=0)=\sqrt{s}\,\Gamma^{\rm vac}_{Y^\ast \to A,B}(s)$.\footnote{Even in medium, in the limit of zero spatial momentum $\mathbf{q}\to \mathbf{0}$ for finite $q^0$, one has the equality $\Pi_\mathrm T(q^0, \mathbf{q})=\Pi_\mathrm L(q^0, \mathbf{q})$ if the corresponding self-energy $\Pi^{\mu\nu}$ obeys the transversality condition $q_\mu \Pi^{\mu\nu}=0$.} In terms of the weights $W_i$, the projected anti-Hermitian self-energies then take the form
\begin{multline}
    \label{eq: anti-Hermitian self-energy vectormediator}
    \Pi^\mathcal A_{{Y,i}}(q)= \frac{\sqrt{s}\, {\Gamma_{Y^\ast\to A,B}^{\rm vac}}(s)}{1+f_Y(q_0)}\int_{-1}^1 \frac{\dd\!\cos \theta^{\rm CM}}{2} W_i(\cos \theta^{\rm CM}) \\ \times[1+\eta_A f_A(E_A)][1+\eta_B f_B(E_B)]\, , {\quad i=\mathrm T,\mathrm L} 
\end{multline}
with the plasma-frame energies $E_A$ and $E_B$ of the bath particles from Eq.~\eqref{eq:Eab_limits}. 

\subsubsection*{Connection to the $2\to2$ Boltzmann collision term}
To make the connection with the standard Boltzmann treatment explicit, we rewrite the CTP production rate in terms of the bath-particle distribution functions and the resummed mediator propagator.

\begin{figure}[t]
    \centering
    \includegraphics[width=0.5\linewidth]{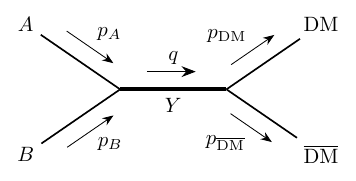}
    \caption{The $2\to2$ annihilation process $A,B\to Y ^\ast \to \text{DM} ,\overline{\text{DM}}$ appearing in the Boltzmann-like form of the CTP production rate for $q_0>0$ and $q^2\geq(m_A+m_B)^2$. The thick line denotes the resummed mediator.}
    \label{fig:only2to2}
\end{figure}

For this purpose, we express $G_{Y,IJ}^<$ entering Eq.~\eqref{eq:ctp-collision-general} as the product
\begin{align}
G_{Y,IJ}^<(q) = G^R_{Y,IK}(q) \, \Pi_{Y}^{<,KL}(q) \, G^A_{Y,LJ}(q) ,
\label{eq:general-resummed-wightman}
\end{align}
as derived in App.~\ref{sec:app:ctp-resummed-propagator}. By substituting the KMS relation $i\Pi_Y^<(q)=2f_Y(q_0)\Pi_Y^\mathcal A(q)$, the one-loop anti-Hermitian self-energy Eq.~\eqref{eq: general antihermiitian selfenergy} and the detailed balance identity for the bosonic mediator,
\begin{align}
\frac{f_Y(q_0)}{1+f_Y(q_0)} \bigl[1 + \eta_A f_A(E_A)\bigr] \bigl[1 + \eta_B f_B(E_B)\bigr] = f_A(E_A)f_B(E_B) ,
\label{eq:detailed-balance-AB}
\end{align}
valid for $q_0=E_A+E_B$, the CTP production rate can be written as
\begin{align}
\gamma_\text{DM} = 2 \int\dd{\Pi}_{A} \dd{\Pi}_{B} {\dd{\Pi}_{\text{DM}} \dd{\Pi}_{\overline{\text{DM}}}} (2\pi)^4 \delta^{(4)}(p_A+p_B-{p_{\text{DM}}-p_{\overline{\text{DM}}} }) f_{A} f_{B} |\mathcal M|^2_{\mathcal T}
\label{eq:temp}
\end{align}
with the effective in-medium squared matrix element
\begin{align}
|\mathcal M|^2_{\mathcal T} = \frac{1}{S_{AB}S_{\rm DM}} \mathcal H_{\rm DM}^{IJ}{(p_{\text{DM}},p_{\overline{\text{DM}}})} \, G^R_{Y,IK}(q) \, \mathcal H_{\rm bath}^{KL}(p_A,p_B) \, G^A_{Y,LJ}(q) .
\end{align}
Hence, for the one-loop bath contribution considered here, the CTP collision term can be cast into the form of a Boltzmann $2\to2$ collision term for $A,B\to Y^\ast \to \text{DM} ,\overline{\text{DM}}$, illustrated in Fig.~\ref{fig:only2to2}.
The thermal modifications are encoded in the retarded and advanced mediator propagators through the corresponding Hermitian and anti-Hermitian self-energies; replacing these by the resummed vacuum Feynman propagator and its complex conjugate recovers the usual vacuum squared matrix element.
For unequal masses, $m_A\neq m_B$, and $q^2\leq(m_A-m_B)^2$, the same mediator self-energy also describes decay-like processes like $A\to B,Y^\ast$. {If the bath particles are gauge charged, the Landau-Pomeranchuk-Migdal (LPM) effect~\cite{Landau:1953gr,Landau:1953um,Migdal:1956tc} corresponding to multiple soft scatterings during the average formation time of the decay $Y\leftrightarrow A,B$ results in another leading-order contribution to $\Pi_Y$ (see, e.g., Ref.~\cite{Arnold:2001ba} for the power-counting scheme). Its computation requires the resummation of ladder diagrams and has only been achieved in the ultra-relativistic limit by neglecting vacuum mass scales~\cite{Anisimov:2010gy,Besak:2012qm,Becker:2025lkc}. Importantly, such corrections enter the CTP production rate for $s$-channel DM exclusively through the mediator and not the DM spectral density.  }

\section{Order-by-Order Analysis of the Dark Matter Production Rate}
\label{sec:order-by-order}

\begin{figure}[t]
    \centering
    \includegraphics[width=1\linewidth]{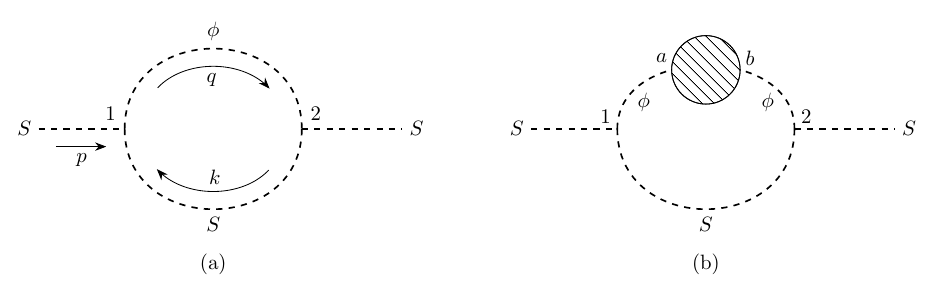}
    \caption{DM self-energy diagrams considered in the order-by-order analysis. Diagram (a) contains a free mediator propagator, while diagram (b) contains a single self-energy insertion in the mediator line. The labels $a$ and $b$ denote internal CTP indices. 
}
\label{fig:scalar-model-diagrams}
\end{figure}

The resummed mediator description in the previous section treats the on-shell and off-shell mediator contributions within a single finite-width spectral density. To identify how the conventional separation into decay and annihilation contributions emerges, we consider the CTP production rate at fixed order in mediator self-energy insertions for a generic interaction $\mathcal L\supset -g_S\,\phi S^2$ of $S$ with a real scalar mediator $\phi$.  The mediator is also coupled to equilibrium bath particles, whose effect is encoded in the mediator self-energy $\Pi_{\phi}^{ab}.$
The DM self-energies with a free scalar mediator and with a single self-energy insertion in the mediator are shown in Fig.~\ref{fig:scalar-model-diagrams}. {In the following, we denote contributions without and with a single mediator self-energy insertion by the superscripts $(0)$ and $(1)$, respectively.} 

The DM self-energy with a free mediator propagator directly reproduces the Boltzmann decay collision term, 
\begin{align}
\gamma_S^{(0)}&=\int \frac{\dd[3]{\mathbf p}}{(2\pi)^3}\frac{\eval{i\Pi_S^<(\omega_p,\mathbf p)}^{(0)}}{2\omega_p}
\nonumber\\
&=4g_S^2\int \dd\Pi_p\int \frac{\dd[4]k}{(2\pi)^4}\frac{\dd[4]q}{(2\pi)^4}(2\pi)^4\delta^{(4)}(q-p-k)
\nonumber\\
&\qquad\times 2\pi\delta(k^2-m_S^2)\bigl[\theta(k^0)+f_S(|k^0|)\bigr]\,2\pi\delta(q^2-m_\phi^2)\bigl[\theta(-q^0)+f_\phi(|q^0|)\bigr]
\nonumber\\ 
&\overset{\mathclap{f_S\ll1}}{\simeq}\int \dd\Pi_p\,\dd\Pi_k\,\dd\Pi_q\,(2\pi)^4\delta^{(4)}(q-p-k)f_\phi(\omega_q)|\mathcal M|^2_{\phi\to S,S}\simeq\mathcal C_{\phi\to S,S}
\label{eq:gammaS_LO}
\end{align}
with the vacuum decay matrix element $|\mathcal M|^2_{\phi\to S, S}=4g_S^2$. The internal DM line is treated in the Kadanoff-Baym ansatz together with the quasi-particle approximation. With $f_S\ll 1$, only its positive-energy contribution $k^0=\omega_k$ remains. Since the external DM momentum satisfies $p^0=\omega_p$, energy conservation implies $q^0>0$ and only the decay  $\phi\to S,S$ remains.
\\

The DM self-energy containing a single self-energy insertion in the mediator line requires the summation over the internal CTP indices $a$ and $b$: 
\begin{align}\label{app:eq: Delta1loopModelA}
    -i\Delta_\phi^{<,(1)}=&\nonumber    \sum_{a,b=1}^2(-1)^{a+b}i\Delta_{0,\phi}^{1a}i\Pi _\phi^{ab}i\Delta_{{0,\phi}}^{b2}
    \\=&2\operatorname{Re}[i\Delta_{0,\phi}^{11}i\Pi^{11}_\phi i\Delta_{0,\phi}^{<}]- i\Delta_{0,\phi}^{11}i\Pi^{<}_\phi i\Delta_{0,\phi}^{22}- i\Delta_{0,\phi}^{<}i\Pi^{>}_\phi i\Delta_{0,\phi}^{<} \, ,
\end{align}
where we have already used that $i\Delta_{0,\phi}^{12}$ is real and $(i\Delta_{0,\phi}^{11})^\ast = i\Delta_{0,\phi}^{22}$ and, analogously, $(i\Pi_\phi^{11})^\ast = i\Pi_\phi^{22}$. 
To show the absence of pathologies, or ``pinch singularities'', in Eq.~\eqref{app:eq: Delta1loopModelA} corresponding to the product of same-argument $\delta$-functions, we use the following functional representations of the Dirac distribution
\begin{align}
\delta(x) = \lim_{\epsilon \to 0^+} \frac{1}{\pi} \frac{\epsilon}{x^2 + \epsilon^2 }\, , \qquad \delta(x) = {\lim_{\epsilon \to 0^+} }\frac{2}{\pi}\frac{\epsilon^{3}}{(x^2+\epsilon^2)^2}   \, .
\end{align}
In addition, we need the principal-value distributions
\begin{align}
     \mathcal{P}\left[\frac{1}{x}\right] = \lim_{\epsilon \to 0^+} \frac{x}{x^2 + \epsilon^2}\, ,  \quad \mathcal{P}\left[\frac{1}{x^2}\right] =  {\lim_{\epsilon \to 0^+}} \frac{x^2-\epsilon^2}{(x^2+\epsilon^2)^2},
\end{align}
where the latter is more precisely understood as the Hadamard finite part. Then, in agreement with Ref.~\cite{Altherr:1994fx}, we find that Eq.~\eqref{app:eq: Delta1loopModelA} is only free of pinch singularities, if the mediator self-energy obeys the KMS relation,
\begin{align}\label{app: kms for cencallation}
f_\text{BE}\operatorname{Re}[i\Pi^{11}_\phi]=\frac{1}{2}(1+2f_\text{BE})i\Pi^{<}_\phi\, .
\end{align}
Lastly, we use that $\mathcal{P}\delta$-terms are not pinch singularities, but finite derivatives of $\delta$-distributions, 
\begin{equation}
     \delta'(x) = -2\,\mathcal{P}\left[\frac{1}{x}\right] \delta(x) \, ,
\end{equation}
such that we arrive at the pathology-free single-insertion contribution to the mediator Wightman propagator:
\begin{align}\label{eq: 1loop-perturbative-prop}
 i\Delta^\text{<,(1)}_\phi(q)= \mathcal{P}\left[\frac{1}{(q^2-m_\phi^2)^2}\right]i\Pi^{<} _\phi(q) -  2\pi \left[\theta(-q^0) + f_\text{BE}(|q_0|)\right]\delta^\prime(q^2-m_\phi^2)\Pi^\mathcal H_\phi(q) \, .
\end{align}
Analogously to the production rate with a free mediator in Eq.~\eqref{eq:gammaS_LO}, inserting $i\Delta_\phi^{<,(1)}$ gives the DM  production rate with a single mediator self-energy insertion,
\begin{align}
    \gamma_S^{(1)}=4g_{S}^2 \int \dd \Pi_p  {\dd\Pi_k } i\Delta_\phi^{<,(1)}(q)\Big|_{q=p+k}\equiv\gamma_{S,\text{off}}^{(1)} + \delta \gamma_{S,\text{pole}}^{(1)}.
\end{align}
Here, the first and second terms of the single-insertion expression in Eq.~\eqref{eq: 1loop-perturbative-prop} are identified with $\gamma_{S,\rm off}^{(1)}$ and $\delta\gamma_{S,\rm pole}^{(1)}$, respectively. The latter describes the fixed-order correction associated with the shift of the mediator pole induced by the Hermitian self-energy, as reflected by the expansion $\delta(q^2-m_\phi^2-\Pi_\phi^{\mathcal H})=\delta(q^2-m_\phi^2)-\Pi_\phi^{\mathcal H}\delta'(q^2-m_\phi^2)+\mathcal O((\Pi_\phi^{\mathcal H})^2)$. The former term is given by 
\begin{align}
    \gamma_{S,\text{off}}^{(1)}= 4g_S^2\int\dd \Pi_p \dd \Pi_k \mathcal P\left[\frac{1}{(q^2-m_\phi^2)^2}\right]i\Pi^<_\phi(q)\Bigg|_{q=p+k}.
\end{align}
At this stage, the mediator self-energy remains unspecified. Specializing $i\Pi_{\phi}^<$ to the same one-loop contribution from equilibrium bath particles $A$ and $B$ as in Sec.~\ref{sec: One-loop bath contribution}, the same phase-space manipulations identify $\gamma_{S,\rm off}^{(1)}$ with the off-shell $2\to2$ contribution $A,B\to\phi^\ast\to S,S$. In contrast to the resummed result of Sec.~\ref{sec: One-loop bath contribution}, the mediator is here given by the fixed-order distribution $\mathcal{P}\!\left[(s-m_\phi^2)^{-2}\right]$, which excludes the resonant on-shell contribution and agrees with the zero-width limit of the PVS and, equivalently, the SDF prescription.
The resulting fixed-order structure agrees with the analysis in Refs.~\cite{Matak:2022qwc,Matak:2023zox} which used holomorphic cutting rules; there, the corresponding dispersive self-energy contribution vanishes under the adopted on-shell renormalization conditions, whereas here $\Pi_\phi^{\mathcal H}$ contains the finite-temperature correction to the mediator pole.

\section{Numerical Examples}
\label{sec: Numerical Examples}

We now compare the DM number-density evolution obtained from the semi-classical Boltzmann framework supplemented by the different RIS-subtraction schemes introduced in Sec.~\ref{sec: The Real Intermediate State Problem} with the CTP result for two benchmark models. We begin with the minimal scalar singlet DM model, followed by a simplified model with a vector mediator. 

In principle, Eq.~\eqref{eq:ctp-collision-general} can account for arbitrary bath-particle production channels proceeding through an $s$-channel process, provided that the mediator self-energy entering the spectral density includes the relevant initial-state contributions. For simplicity, we restrict the numerical analysis to one-loop bath self-energies. For the equal-mass initial states considered in the benchmark models, the one-loop bath self-energy describes bath-particle annihilation and therefore gives rise to $2\to2$ production processes, as discussed in Sec.~\ref{sec: One-loop bath contribution}. For a consistent comparison, in contrast to the usual procedure where the width in the Breit--Wigner propagator is evaluated on shell, the mediator's width is momentum-dependent in both, the CTP and Boltzmann collision term, and evaluated at the same perturbative order. We denote the corresponding invariant-mass-dependent vacuum width by $\Gamma_Y^{\rm vac}(s)$, while $\Gamma_Y^{\rm vac}\equiv\Gamma_Y^{\rm vac}(m_Y^2)$ denotes its on-shell value.

\paragraph{Numerical setup} For the numerical computation of the relic abundance, we apply the Friedmann equation $H^2=\rho/(3 M^2_{\rm Pl})$ for a flat Universe with the reduced Planck mass $M_{\rm Pl}$ to express the Hubble rate in terms of the total energy density of the Universe $\rho$ and rewrite the Boltzmann Eq.~\eqref{Eq: BoltzmannEq} {as well as the CTP number-density evolution Eqs.~\eqref{eq: scalar DM evolution equation} and \eqref{eqs: fermionic and vector dm}}
in terms of the photon temperature $T$ via the dimensionless quantity $x = m_{\mathrm{DM}}/T$. Further, assuming entropy conservation $\dv{t}({\mathbb{s}} a^3)=0$, the yield $Y_{\mathrm{DM}} = n_{\mathrm{DM}}/{\mathbb{s}}$ is introduced as a comoving number density. Given the freeze-in condition, the contribution of DM to $\rho$ and {the total entropy density $\mathbb{s}$} is assumed to be negligible, such that we express both in terms of the tabulated~\cite{Drees_2015-relativisticDOF} Standard Model effective numbers of degrees of freedom, $h_{\mathrm{eff}}$ and $g_{\mathrm{eff}}$, via $\rho=g_{\rm eff}(T)\frac{\pi^2}{30}T^4$ and ${\mathbb{s}}=h_{\rm eff}(T)\frac{2\pi^2}{45}T^3$. For the numerical analysis, we use $\gamma_\text{DM}$ as a common notation for the DM collision term, corresponding to $\mathcal C_\text{DM}$ in the Boltzmann formalism and to the production rate defined in Sec.~\ref{sec: Evolution Equation for the Dark Matter Density} in the CTP framework. Numerical integrations are performed using the GNU Scientific Library (GSL)~\cite{Galassi:2019czg}.
We then solve 
\begin{align}
    \dv{Y_{\mathrm{DM}}}{x}=\frac{\mathcal{\gamma}_{\mathrm{DM}}}{{\mathbb{s}} \overline{H} x},
\end{align}
down to a temperature $T_{f}$ at which further DM production can be safely neglected, and obtain today's DM abundance 
\begin{align}\label{eq: relicdensity}
   \Omega_{\rm DM} h^2= \frac{m_{\rm DM}\,  n_{\rm DM}(T_f)}{\rho_c/h^2}\frac{h_{\rm eff}(T_0)}{h_{\rm eff}(T_f)}\frac{T_0^3}{T_f^3}
\end{align}
solely by taking into account cosmic expansion through the present-day photon temperature $T_0$ where $\rho_c$ is the critical density of the Universe. For the \texttt{micrOMEGAs} reference results, we use version \texttt{7.1.4}~\cite{Belanger:2026asz} and compute the freeze-in relic density with the \texttt{darkOmegaFi} routine.

\paragraph{Spectral kernels and Boltzmann prescriptions}

For a more direct analytical and numerical comparison, we write the Boltzmann collision terms for the different subtraction schemes in the same form as the CTP collision term discussed in Sec.~\ref{sec:CTP-Resummed}, namely as integrals over a spectral quantity. In the Boltzmann case, we refer to these quantities as \emph{spectral kernels}, since they do not in general define normalized spectral densities. With our conventions, a bosonic spectral density satisfies the sum rule \cite{LeBellac_1996}
\begin{align}
\label{eq:spectral_sum_rule}
\int_0^\infty \frac{\dd{q}^0}{\pi} \, q^0 \, \rho(q^0,\mathbf q) = 1 .
\end{align}
For later convenience, we collect the spectral densities and kernels entering the different treatments of the mediator resonance. For a bosonic mediator $Y$ with four-momentum $q$, we write $s = q^2$.
The finite-width CTP spectral density is given by Eq.~\eqref{eq:scalar-spectral-density}. In the zero-width limit, and neglecting the mass shift induced by the Hermitian self-energy, the scalar spectral density reduces to
\begin{align}
\rho_{\rm dec}(q) = 2\Delta_0^{\mathcal A}(q) = 2\pi \, \delta(s-m_Y^2), \qquad q^0 > 0 .
\label{eq:rho_zero_width}
\end{align}
Both the finite-width spectral density Eq.~\eqref{eq:scalar-spectral-density} and its zero-width limit~\eqref{eq:rho_zero_width} are normalized according to the spectral sum rule Eq.~\eqref{eq:spectral_sum_rule}.

On the other hand, as discussed in Sec.~\ref{sec: One-loop bath contribution}, the conventional Boltzmann $2\to2$ collision term is obtained when thermal corrections to the mediator width and the Hermitian part of the self-energy are neglected in the denominator of the propagator. We therefore define the corresponding spectral kernel as 

\begin{align}
\rho_{2\to2}^{\rm BW}(q) = \frac{2\Pi_{{Y}}^{\mathcal A}(q,T)}{(s-m_Y^2)^2 + [\Pi_{{Y}}^{\mathcal A}(s,T=0)]^2},
\end{align}
where the denominator is the familiar vacuum Breit--Wigner form, since $\Pi_{{Y}}^{\mathcal A}(s,T=0)=\sqrt s \Gamma_Y^\text{vac}(s)$ . 

The different Boltzmann prescriptions introduced in Sec.~\ref{sec: The Real Intermediate State Problem} and used {within the numerical comparison} can then be written as
\begin{subequations}
\label{eq:rho_boltzmann_schemes}
\begin{align}
\rho_{\rm B}^{\rm real}(q)&=\rho_{\rm dec}(q)+2\Pi^{\mathcal A}_Y(q,T)\frac{(s-m_Y^2)^2}{\left[(s-m_Y^2)^2 + [\Pi_{{Y}}^{\mathcal A}(s,T=0)]^2\right]^2},
\\[1ex]
\rho_{\rm B}^{\rm PVS}(q)&=\rho_{\rm dec}(q)+2\Pi_{{Y}}^{\mathcal A}(q,T)\frac{(s-m_Y^2)^2 - [\Pi_{{Y}}^{\mathcal A}(s,T=0)]^2}{\left[(s-m_Y^2)^2 + [\Pi_{{Y}}^{\mathcal A}(s,T=0)]^2\right]^2}, 
\label{eq: spectral kernel PVS}
\\[1ex]
\rho_{\rm B}^{\rm SDF}(q)&=\rho_{\rm dec}(q)+\rho_{2\to2}^{\rm BW}(q)-2\pi \, \frac{\Pi_{{Y}}^{\mathcal A}(q,T)}{\Pi_{{Y}}^{\mathcal A}(m_Y^2,T=0)} \, \delta(s-m_Y^2).
\end{align}
\end{subequations}
The cut scheme can be written in the same notation as 
\begin{align}\label{eq: rhocutscheme}
\rho_{\rm B}^{\rm Cut}(q) = R(\delta) \, \rho_{\rm dec}(q) + [1-\Theta\big(s-m_Y^2, \delta(s)\big)] \, \rho_{2\to2}^{\rm BW}(q),
\end{align}
with the weight function
\begin{align}
R(\delta) = \frac{1}{\pi} \int \dd s \, {\Theta}\big(s-m_Y^2,\delta(s)\big) \, \frac{\Pi_{{Y}}^{\mathcal A}(s,T=0)}{(s-m_Y^2)^2 + [\Pi_{{Y}}^{\mathcal A}(s,T=0)]^2} \,
\end{align}
that accounts for the part of the resonant contribution retained after applying the cut on the invariant mass. The separation scale $\delta$ is in principle a free parameter of the cut scheme. Throughout the numerical analysis, we choose it to follow the momentum-dependent vacuum width, $\delta(s)=\sqrt s \, \Gamma_Y^{\rm vac}(s)=\Pi_Y^\mathcal A (s, T=0)$.

\subsection{Scalar Singlet Dark Matter}
\label{sec: Scalar Singlet Dark Matter}

As a first model, we consider the scalar singlet DM model \cite{Silveira:1985rk,scalarsingletIntro1-McDonald_1994, Burgess_2001-scalarsngletintroduction2}, a minimal extension of the SM in which the dark sector couples to the visible sector through the Higgs portal. Due to its minimality, this model has been extensively studied, specifically in the context of freeze-out \cite{Cline_2013-WIMP-SSDM, 2019-WIMP-SSDM,Arcadi_2018-WIMP-SSDM,Binder:2017rgn,Ala-Mattinen:2019mpa,Laine:2022ner,Biondini:2025ihi, EscuderoAbenza:2025cfj}. In particular, Ref.~\cite{Laine:2022ner} studied resonant scalar singlet DM annihilation at NLO within a fixed-order treatment. 
It has also been studied as a FIMP candidate in the freeze-in scenario, first in \cite{Yaguna_2011-scalarsingletFIMP} and, e.g., additionally in \cite{Lebedev:2019ton, Bernal:2018kcw, Bringmann_2022, Arcadi:2024wwg}.\\

The model introduces a real scalar field $S$, which is a singlet under the SM gauge group and serves as the DM candidate. 
It transforms odd under a $\mathbb Z_2$ symmetry, ensuring its stability.  
The Lagrangian before electroweak symmetry breaking, expressed in terms of the SM Lagrangian $\mathcal{L}_{\mathrm{SM}}$ and the Higgs doublet $H$, is given by
\begin{align}
    \mathcal{L}= \mathcal{L}_{\mathrm{SM}}+\frac{1}{2}\,\partial_\mu S\,\partial^\mu S- \frac{m_0^2}{2}\,S^2- \frac{\lambda_S}{4}\,S^4- \frac{1}{2}\lambda\,S^2 H^\dagger H \, .
\end{align}
After electroweak symmetry breaking, with $H=(0,\,v+h)^T/\sqrt{2}$, the DM mass is given by $m_S = \sqrt{m_0^2+ \lambda  v^2/2}$.
The resulting interactions are
\begin{align}
    \mathcal{L} \supset - \frac{1}{2}\lambda v h S^2 - \frac{\lambda}{4} h^2 S^2 .
\end{align}
The parameters relevant for the freeze-in production considered here are therefore the scalar mass $m_S$ and the Higgs-portal coupling $\lambda$. The self-coupling $\lambda_S$ is irrelevant here, although dark-sector self-interactions can lead to cannibal or self-populating freeze-ins~\cite{March-Russell:2020nun,Cervantes:2024ipg}.

The production of DM after electroweak symmetry breaking is mediated by the SM Higgs boson. 
The relevant production processes are SM annihilation processes into an $S$-pair through an $s$-channel Higgs propagator and, whenever kinematically allowed, Higgs decays, $h \to SS$. 
Close to the Higgs resonance, the resonant part of the annihilation contribution is directly related to the decay contribution and therefore provides a simple setting for studying the RIS double-counting problem.
 
While in large parts of the parameter space production through $W^+W^-$ annihilation dominates \cite{Yaguna_2011-scalarsingletFIMP}, in the parameter region relevant for this work, close to the Higgs resonance, $b\bar b$-annihilation gives the dominant contribution. 
Since our focus is the RIS treatment rather than a precision prediction, we retain only the $b\bar{b}$-channel, using the four-loop $\overline{\mathrm{MS}}$ bottom mass, as implemented in \texttt{RunDec}~\cite{Chetyrkin:2000yt}, evaluated at the scale $\mu =2m_S$. Furthermore, we consider only production after electroweak symmetry breaking, which is the dominant contribution in the parameter space considered~\cite{Bringmann_2022}. 
We therefore impose the initial condition \(Y_{S}(T_{\rm EW})=0\), taking \(T_{\rm EW}=\SI{160}{\giga\electronvolt}\) as the electroweak crossover temperature, consistent with lattice determinations \cite{D_Onofrio_2016-EWSBTemp}.\footnote{Throughout our example models, we remain agnostic about inflationary particle production since, e.g., $S$ as a GeV scale feebly interacting stable scalar with $T_{\rm EW}$ as reheating temperature might already be overproduced by classical gravity during (high scale) inflation~\cite{Lebedev:2022cic,Feiteira:2025rpe}.}

\subsubsection*{CTP and Boltzmann collision terms}

For the scalar singlet model, the off-shell partial decay width into DM is given by
\begin{align}
   \Gamma^{\rm vac}_{h^\ast \to S,S}(s)=\frac{\lambda^2 v^2 }{32 \pi \sqrt s}\, \beta_S , \qquad \beta_S=\sqrt{1-\frac{4m_S^2}{s}}.
\end{align}
Using $i\Delta_h^<(q)=f_\text{BE}(q_0)\rho_h(q)$, the isotropy of the thermal bath, and the substitution $q_0=\sqrt{s+|\mathbf{q}|^2}$, Eq.~\eqref{eq: general collsionterm CTP scalar} can be recast as
\begin{align}
\gamma_\text{DM}[\rho_h]=\frac{\lambda^2 v^2}{64\pi^4}\int_0^\infty \dd |\mathbf q|\,|\mathbf q|^2\int_{4m_S^2}^\infty \dd s\,\frac{\beta_S}{q_0}\,f_\text{BE}(q_0)\,\rho_h(q),
\end{align}
The CTP and Boltzmann treatments are encoded in the choice of $\rho_h$.

On the bath side, we retain the bottom-quark contribution to the Higgs self-energy. The corresponding absorptive self-energy $\Pi^\mathcal A_h(q,T)$ entering $\rho_h$ is given by Eq.~\eqref{eq:scalar-thermal-PiA}, with the vacuum width
\begin{align}
    \Gamma^\text{vac}_{h^\ast\to b ,\bar b}(s)=\frac{N_c^b y_b^2}{16\pi}\sqrt s \,\left(1-\frac{4m_b^2}{s}\right)^{3/2},
\end{align}
where $N^b_c=3$ denotes the number of colors of the bottom quark. Since this is the only bath contribution retained here, we identify $\Gamma_h^{\rm vac}(s)=\Gamma_{h^\ast\to b\bar b}^{\rm vac}(s)$. The corresponding Hermitian self-energy from the bottom-quark loop is
\begin{align}
\Pi_h^\mathcal H(q,T)\hspace{-0.1em}=\hspace{-0.1em}&\frac{y_b^2}{2} \frac{N^b_c}{|\mathbf q|} \hspace{-0.1em}\int_{m_{b}}^\infty \hspace{-0.1em}\frac{\dd\omega_k}{4\pi^2}\Bigg[{(4m_{b}^2-q^2)}\ln{\left|\frac{(q^2+2|\mathbf k||\mathbf q|)^2-4\omega_k^2q_0^2}{(q^2-2|\mathbf k||\mathbf q|)^2-4\omega_k^2q_0^2}\right|}+8|\mathbf k| |\mathbf q|\Bigg]f_\text{FD}(\omega_k),
\label{eq:Pihbbar}
\end{align}
with $|\mathbf k|= \sqrt{\omega_k^2-m_b^2}$. We also consider the Hard Thermal Loop (HTL) approximation, given by
\begin{align}
    \Pi_{h,\mathrm{HTL}}^{\mathcal H}(T)=\frac{y_b^2 N_c^b T^2}{12}\, .
    \label{eq:PihbbarHTL}
\end{align}
A detailed derivation of the Hermitian self-energy, together with a discussion and numerical comparison of the HTL approximation, is given in App.~\ref{sec: app: hermitian self-energys}.
We stress that Eqs.~\eqref{eq:Pihbbar} and \eqref{eq:Pihbbar} only contain the bottom-quark loop since we want to illustrate our RIS subtraction procedure. For a precision study of the singlet scalar freeze-in abundance as in Ref.~\cite{Bringmann_2022}, the full Standard Model has to be taken into account, as the thermal Higgs mass is dominated by gauge, top-Yukawa and Higgs self-interactions~\cite{Carrington:1991hz,Quiros:1999jp,Curtin:2016urg}.  

\subsubsection*{Relic-density comparison}

\begin{figure}[tbp]
    \centering

    \begin{subfigure}[b]{0.49\textwidth}
        \centering
        \includegraphics[width=\textwidth]{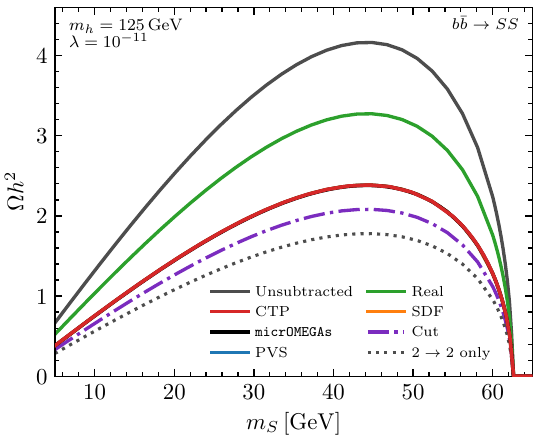}
        \caption{}
        \label{fig:ModelA-massscan-new}
    \end{subfigure}
    \hfill
    \begin{subfigure}[b]{0.49\textwidth}
        \centering
        \includegraphics[width=\textwidth]{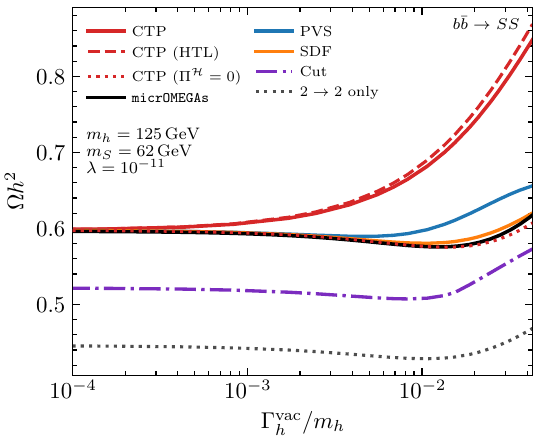}
        \caption{}
        \label{fig:modelA-widthscan-new}
    \end{subfigure}

     \caption{Comparison of the relic density calculated within the Boltzmann formalism for different subtraction schemes with the result obtained in the CTP formalism for the scalar singlet DM model. Panel (a) shows the relic density as a function of the DM mass. For the narrow Higgs width considered, the SDF, PVS and \texttt{micrOMEGAs} results are not separately visible, as they overlap with the CTP curve. Panel (b) shows the relic density at fixed DM mass as a function of the normalized on-shell vacuum  width.}
    \label{fig:modelA-relic-density-mass-and-width-scan}
\end{figure}

Figure~\ref{fig:ModelA-massscan-new} shows the scalar singlet relic density for different masses of $S$. For $m_S<m_h/2$, the decay $h\to S,S$ is kinematically allowed and the Higgs can be produced on shell. The unsubtracted result {(gray)} adds naively the decay and annihilation Boltzmann collision terms and therefore double counts the real intermediate state. Away from the threshold region, this leads to an overestimate of the CTP relic density by approximately $\SI{75}{\percent}$. This ratio is less than two since the Pauli-blocking factors contained in the anti-Hermitian mediator self-energy through $\bar G_{b\bar b}$ reduce the effective mediator width compared to its vacuum value, $\Pi_h^{\mathcal A}(q,T)/\Pi_h^{\mathcal A}(q,0)<1$, thereby enhancing the CTP production rate over the $2\to2$ Boltzmann collision term. Finite-temperature statistical effects suppress the resonant $2\to2$ contribution relative to the decay contribution, as reflected by $\Pi_h^{\mathcal A}(q,T)/\Pi_h^{\mathcal A}(q,0)<1$, with Pauli blocking entering through $\bar G_{b\bar b}$ and an additional suppression arising from the factor $1/[1+f_h(q_0)]$.  Consequently, the unsubtracted result is enhanced by less than a factor of two. As shown in Eq.~\eqref{eq: realproof}, the real-part prescription {(green)} removes only half of the resonant contribution in the zero-width limit and therefore still overestimates the CTP result by $\sim \SI{37}{\percent}$ for these benchmark parameters. Conversely, retaining only the unsubtracted $2\to2$ contribution {(dotted)} while omitting the decay term underestimates the result by approximately $\SI{25}{\percent}$. For the cut scheme (purple dash-dotted), the arbitrary separation between the resonant and off-shell regions is set by the invariant-mass window $\delta$. With the choice $\delta=\sqrt s \Gamma_h^{\rm vac}(s)$ used here, the resulting relic density lies below the CTP result. For the narrow mediator width present in this model, the CTP {(red)}, PVS {(blue)}, SDF {(orange)} and \texttt{micrOMEGAs} results {(black)} are nearly indistinguishable. As $m_S$ approaches $m_h/2$, the phase space for the on-shell decay becomes suppressed. Above the threshold, neither the decay nor an on-shell RIS contribution is kinematically accessible, such that the resonant enhancement is absent and production proceeds through the off-shell tail of the mediator. {At $m_S=\SI{62}{\giga\electronvolt}$, the on-shell width is $\Gamma_h^{\rm vac}/m_h \simeq 1.5\times10^{-5}$. To study deviations from the narrow-width limit, we fix $m_S$ in Fig.~\ref{fig:modelA-widthscan-new} and artificially increase the bottom-quark Yukawa coupling while keeping $m_b$ constant, thereby varying $\Gamma_h^{\rm vac}$.}
At small widths, the CTP, PVS, and SDF results are nearly identical. At larger widths, however, PVS and SDF begin to separate. At the largest width shown, $\Gamma^\text{vac}_h/m_h\simeq \SI{4.3}{\percent}$, their relic densities differ by approximately $\SI{6}{\percent}$. The Hermitian part of the mediator self-energy also becomes increasingly relevant with increasing width. For this near-threshold benchmark with $m_S=\SI{62}{\giga\electronvolt}$, the corresponding shift of the spectral peak towards larger invariant masses increases the available final-state phase space and enhances the relic density. At the largest width shown, the full CTP result is approximately $\SI{40}{\percent}$ larger than the result obtained with $\Pi^{\mathcal H}_h=0$ {(red dotted)}. {Since the dominant SM contributions to the Higgs thermal mass are not included in $\Pi^{\mathcal{H}}_h$, this sensitivity should be read as an indication that a complete thermal Higgs self-energy is required for precise predictions near the threshold.} The \texttt{micrOMEGAs} result closely follows the latter over the entire width scan and differs from it by approximately $\SI{2}{\percent}$ at the largest width. The increasing separation from the full CTP curve therefore arises predominantly from the Hermitian self-energy and the associated thermal mass shift.
The HTL approximation (red dashed) closely follows the full CTP result in the narrow-width regime. Its relative deviation reaches the percent level at $\Gamma_h^{\rm vac}/m_h\gtrsim 10^{-2}$ and subsequently grows monotonically, with the HTL result exceeding the full CTP result by approximately $\SI{2.4}{\percent}$ at the maximal width considered.

\subsection{Vector-Mediated Scalar Dark Matter}
\label{sec:vetor-model}

To extend the numerical study from scalar to vector mediators, we consider a simplified model with a massive spin-$1$ mediator $Z'_\mu$ and a complex scalar singlet $S$ as DM. Such a setup has, e.g., been studied in the freeze-out scenario in Ref.~\cite{Chang:2022jgo}. Earlier works on vector-mediated freeze-in assuming high as well as low reheating temperatures include, e.g., Refs.~\cite{Belanger:2018ccd,Cosme:2021baj,Arcadi:2024obp}. As in the scalar singlet model, the DM stability is ensured by a $\mathbb{Z}_2$ symmetry under which $S$ is odd, while the SM fields and the mediator are even.
The model Lagrangian is given by 
\begin{align}
\mathcal L=\mathcal{L}_{\rm SM}&+\partial_\mu S^\dagger \partial^\mu S-m_ S^2 S^\dagger S-\frac{1}{4} F'_{\mu\nu}F'^{\mu\nu}+\frac{1}{2} m_{Z'}^2 Z'_\mu Z'^\mu
\nonumber \\
&+g_q Z'_\mu \sum_q \bar{q}\gamma^\mu q+i g_{\rm DM} Z'_\mu\left[S^\dagger(\partial^\mu S)-(\partial^\mu S^\dagger)S\right],
\end{align}
where $F'_{\mu\nu}=\partial_\mu Z'_\nu-\partial_\nu Z'_\mu$ denotes the mediator's field-strength tensor.
The model is characterized by four free parameters,
\begin{align}
m_{ S},\, m_{Z'},\, g_q,\, g_{\rm DM},
\end{align}
corresponding to the DM and mediator masses, as well as the mediator couplings to quarks and DM, respectively.

In contrast to the scalar singlet model considered before, we include all six quark flavors in the thermal bath. Since the characteristic energies in the parameter region considered are much larger than the quark masses, the six flavors give approximately equal contributions. In the numerical implementation, the quark masses are set to zero above $T_{\rm EW}=\SI{160}{\giga\electronvolt}$, while their finite masses are retained below this temperature. We choose a reheating temperature of $T_{\rm RH}= \SI{e6}{\giga\electronvolt}$.

\subsubsection*{CTP and Boltzmann collision terms}

For the complex scalar DM candidate, the off-shell partial decay width of the vector mediator is
\begin{align}
    \Gamma^{\rm vac}_{Z'^*\to S, S^\dagger }(s)=\frac{g_\text{DM}^2 \sqrt s}{48 \pi}\beta_S^3, \qquad \beta_S=\sqrt{1-\frac{4m_S^2}{s}}.
\end{align}
In the following, we suppress the mediator label $Z'$ for brevity and write $\Pi_{\mathrm T,\mathrm L}$ and $\rho_{\mathrm T,\mathrm L}$.
Applying the general vector-mediator expression provided in Eq.~\eqref{eq_ general collsion term vectormediator}, the collision term may be expressed as
\begin{align}
\gamma_{\rm DM}[\rho_ \mathrm T,\rho_\mathrm L]=\frac{g_{\rm DM}^2}{96\pi^4}\int_0^\infty \dd |\mathbf q|\,|\mathbf q|^2\int_{4m_S^2}^\infty \dd s\,\frac{s\beta_S^3}{q_0}\,f_\text{BE}(q_0)\left[2\rho_\mathrm T(q)+\rho_\mathrm L(q)\right].
\end{align}
 The CTP collision term involves the corresponding spectral densities $\rho_\mathrm T$ and $\rho_\mathrm L$ of Eq.~\eqref{eq: spectral density vectormediator}, which are constructed from the projected absorptive and Hermitian self-energies $\Pi_i^{\mathcal A}(q,T)$ and $\Pi_i^{\mathcal H}(q,T)$, with $i=\mathrm T,\mathrm L$. The Boltzmann spectral kernels introduced at the beginning of this section are applied analogously to each polarization mode using the corresponding $\Pi_i^{\mathcal A}(q,T)$.

For each quark flavor $q$, the absorptive self-energy components $\Pi_{i,q}^{\mathcal A}(q,T)$ are obtained from Eq.~\eqref{eq: anti-Hermitian self-energy vectormediator} using the vacuum width
\begin{align}
\Gamma_{Z'^\ast\to q,\bar q}^{\rm vac}(s) = \frac{N^q_c g_q^2\sqrt{s}}{12\pi}\left(1+\frac{2m_q^2}{s}\right)\beta_q,
\qquad
\beta_q=\sqrt{1-\frac{4m_q^2}{s}},
\end{align}
together with the separate weight functions for the transverse and longitudinal modes,
\begin{subequations}
\begin{align}
W_\mathrm T^{(q)}(\cos\theta^{\text{CM}}) &= \frac{3\left[2 - \beta_q^2 + \beta_q^2\cos^2\theta^{\text{CM}}\right]}{2\left(3 - \beta_q^2\right)}, \\ W_\mathrm L^{(q)}(\cos\theta^{\text{CM}}) &= \frac{3\left[1 - \beta_q^2\cos^2\theta^{\text{CM}}\right]}{3 - \beta_q^2}.
\end{align}
\end{subequations}
The total absorptive self-energies are obtained by summing the individual quark contributions. In vacuum, the two polarization modes become degenerate, $\Pi_\mathrm T^{\mathcal A}(s,0)=\Pi_\mathrm L^{\mathcal A}(s,0)=\sqrt{s}\,\Gamma_{Z'}^{\rm vac}(s)$, with $\Gamma_{Z'}^{\rm vac}(s)=\sum_q\Gamma_{{Z'}^\ast\to q,\bar q}^{\rm vac}(s)$, such that the same vacuum absorptive self-energy enters the Breit--Wigner denominators for both modes.

The full one-loop expressions for the corresponding transverse and longitudinal Hermitian self-energies $\Pi_{\mathrm T,\mathrm L}^{\mathcal H}(q,T)$, adapted from Ref.~\cite{Drewes:2024wbw-Neff3}, are given in App.~\ref{sec: app: hermitian self-energys}. Their HTL expressions are
\begin{subequations}\label{eq:HTL vector}
\begin{align}
\Pi_{\mathrm L,\mathrm{HTL}}^{\mathcal H}(q,T) &= \sum_q \frac{N_c^q g_q^2 T^2}{3} \left(1-\frac{q_0^2}{|\mathbf q|^2}\right) \left[ 1-\frac{q_0}{2|\mathbf q|} \ln\left| \frac{q_0+|\mathbf q|}{q_0-|\mathbf q|} \right| \right], \\
\Pi_{\mathrm T,\mathrm{HTL}}^{\mathcal H}(q,T) &= \sum_q \frac{N_c^q g_q^2 T^2}{6} \frac{q_0^2}{|\mathbf q|^2} \left[ 1- \left(1-\frac{|\mathbf q|^2}{q_0^2}\right) \frac{q_0}{2|\mathbf q|} \ln\left| \frac{q_0+|\mathbf q|}{q_0-|\mathbf q|} \right| \right].
\end{align}
\end{subequations}

\subsubsection*{Relic-density comparison}

\begin{figure}[tbp]
    \centering

    \begin{subfigure}[b]{0.49\textwidth}
        \centering
        \includegraphics[width=\textwidth]{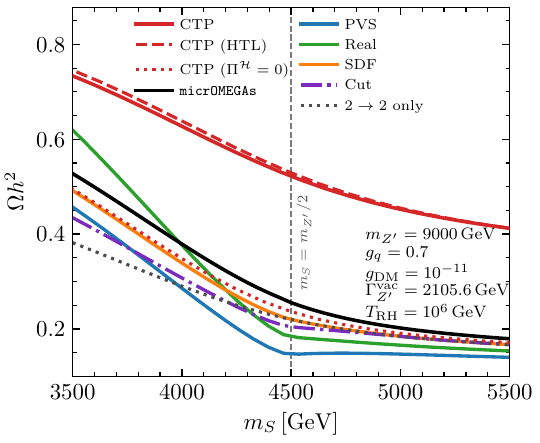}
        \caption{}
        \label{fig:modelC-massscan}
    \end{subfigure}
    \hfill
    \begin{subfigure}[b]{0.49\textwidth}
        \centering
        \includegraphics[width=\textwidth]{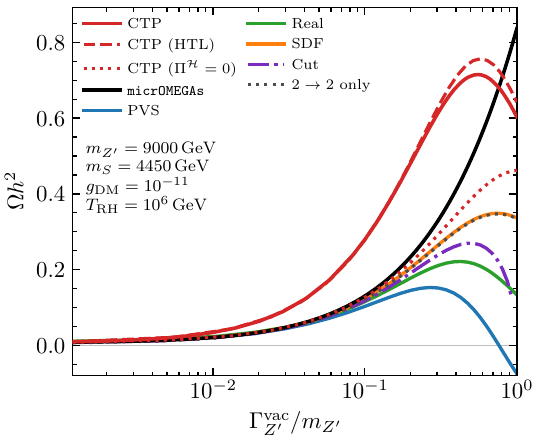}
        \caption{}
        \label{fig:modelC-gqscan}
    \end{subfigure}
    
     \caption{Comparison of the relic density for the vector mediator model calculated within the Boltzmann formalism for different subtraction schemes with the corresponding CTP result. Panel (a) shows the relic density as a function of the DM mass $m_S$ at fixed mediator couplings, thereby probing different kinematic configurations relative to the mediator resonance. The vertical dashed line marks the threshold $m_S=m_{Z'}/2$, above which the on-shell decay $Z'\to S^\dagger, S$ is kinematically forbidden. Panel (b) shows the relic density at fixed DM mass as a function of the normalized on-shell vacuum width $\Gamma_{Z'}^\text{vac}/m_{Z'}$, varied through the mediator--quark coupling $g_q.$ }
    \label{fig:modelC-relic-density}
\end{figure}
In Fig.~\ref{fig:modelC-massscan}, the DM mass $m_S$ is varied for fixed coupling $g_q=\SI{0.7}{}$. As expected, the resulting relic density depends again sensitively on the distance to the threshold $m_S= m_{Z'}/2$. 
While the SDF (orange) and CTP relic densities with the Hermitian mediator self-energy set to zero (red dotted) are in good agreement, the PVS result (blue) underestimates this time the relic density. This is due to the sizeable mediator width in this benchmark. The thermal mass shift induced by the Hermitian mediator self-energy increases the abundance by approximately $50\%$ at the lower end of the displayed mass range, with an even larger enhancement closer to and above the threshold. The HTL result (red dashed) reproduces the full CTP computation (red solid) with percent-level accuracy. The \texttt{micrOMEGAs} result (black) remains close to the CTP result obtained with $\Pi^\mathcal H_{\mathrm T,\mathrm L}=0$, while it differs substantially from the full CTP result.
By increasing $m_S$, the kinematically accessible invariant masses move progressively away from the mediator pole, such that the virtuality $(s-m_{Z'}^2)^2$ increasingly dominates over the width-dependent terms. The differences between the corresponding off-shell kernels therefore become suppressed, and the curves approach each other. 

In Fig.~\ref{fig:modelC-gqscan}, {the mediator width is varied via the coupling $g_q$ while all other parameters remain fixed}, thereby probing the sensitivity of the different prescriptions to increasingly broad resonances. At the smallest width shown, $\Gamma_{Z'}^\text{vac}/m_{Z'}\simeq 10^{-3}$, \texttt{micrOMEGAs} nearly coincides with the CTP result obtained with $\Pi_{\mathrm T,\mathrm L}^\mathcal H=0$, while PVS and SDF lie only about one percent above it. The full and HTL CTP results are already about $15\%$ larger due to the effect of the Hermitian self-energy. With increasing width, the separation between the different predictions becomes more pronounced. For sufficiently large width, the PVS scheme even yields an unphysical negative relic density. This behavior originates from the negative region of the PVS kernel in Eq.~\eqref{eq: spectral kernel PVS}, which can lead to an over-subtraction of the resonant contribution. In contrast, the cut scheme (purple dash-dotted) cannot suffer from this problem: by construction, its spectral kernel contains only non-negative contributions, as the resonant region is excluded rather than subtracted.

The Hermitian self-energy also becomes increasingly relevant for broader resonances. Its associated thermal mass shift changes the position of the spectral weight relative to the nearby DM threshold, leading to a visible separation between the full CTP result and the result obtained with $\Pi_{\mathrm T,\mathrm L}^\mathcal H=0$. The HTL approximation agrees with the full CTP result at the percent level up to $\Gamma_{Z'}^{\rm vac}/m_{Z'}\simeq 0.2$, indicating that the dominant thermal mass correction is well captured by the HTL approximation in this regime. For broader resonances, the difference increases, reaching approximately $6.5\%$ at the largest width shown. The \texttt{micrOMEGAs} result agrees with the CTP result obtained with $\Pi_{\mathrm T,\mathrm L}^\mathcal H=0$ to within approximately $1\%$ for $\Gamma_{Z'}^{\rm vac}/m_{Z'}\lesssim 0.04$. For broader resonances, the two curves separate since \texttt{micrOMEGAs} uses a momentum-independent on-shell width, whereas the CTP result retains the full momentum dependence of $\Pi^{\mathcal A}$. The extension to very large values of $\Gamma_{Z'}^{\rm vac}/m_{Z'}$ is intended primarily as a schematic illustration of the behavior of the different prescriptions beyond the narrow-width regime. As the width becomes comparable to the mediator mass, the separation between an on-shell intermediate state and the off-shell part of the propagator becomes increasingly ambiguous.

\section{Conclusion}\label{sec:Conclusion}

We investigated the treatment of real intermediate states in $s$-channel freeze-in production of DM with an unstable mediator, where the naive simultaneous inclusion of decay and Breit--Wigner-regulated annihilation collision terms within the semi-classical Boltzmann approach leads to a double counting of on-shell mediators. Instead of relying on an ad-hoc RIS-subtraction scheme, we started systematically from the Kadanoff--Baym equations of non-equilibrium QFT. Performing the standard Wigner transformation and gradient expansion as well as exploiting the freeze-in condition $f_{\rm DM}\ll1$, we derived local evolution equations for the number density of a single scalar, fermion or vector DM species. In this description, the DM production rate is given by the lesser DM Wightman self-energy projected onto the physical on-shell states such that, besides being intrinsically free of non-unitary overcounting, quantum statistics and in-medium effects known to be relevant for freeze-in are systematically included.

For $s$-channel pair production via a single scalar or vector mediator in thermal equilibrium with the surrounding bath, we then expressed the production rate in terms of the 1PI-resummed mediator spectral density, which contains the resonant and the off-shell contributions within a single expression and is determined by the Hermitian and anti-Hermitian mediator self-energies. For a vector mediator coupled to a conserved current, the transverse and longitudinal polarization modes thereby acquire in general separate thermal widths and mass shifts. We showed that, for a one-loop mediator self-energy generated by bath particles, the CTP production rate can be cast into the form of a Boltzmann $2\to2$ collision term with an in-medium squared matrix element in which the vacuum Breit--Wigner width is replaced by the full momentum-dependent anti-Hermitian self-energy. For a scalar mediator, and after neglecting the thermal mass shift induced by the Hermitian self-energy, this reproduces the effective width recommended by \texttt{DarkSUSY}~\cite{Bringmann:2018lay,Bringmann_2022}, which is thereby put on a first-principles footing. By re-expanding the mediator propagator order-by-order in self-energy insertions, we furthermore showed that the contribution with a free mediator propagator corresponds exactly to the Boltzmann decay collision term, while for a single mediator self-energy insertion
 the thermal pinch singularities and all on-shell $\delta^2$-contributions cancel, leaving a finite contribution describing $s$-channel production processes, such as $2\to2$ annihilation, through the principal value, or more precisely the Hadamard finite part, of the squared zero-width propagator and a thermal mass correction. The fixed-order result is thus non-singular and identifies in the zero-width limit the PVS and, equivalently, the SDF prescription as the consistent RIS subtraction, in agreement with the holomorphic-cut analysis of Refs.~\cite{Matak:2022qwc,Matak:2023zox}.

The phenomenological relevance of a consistent RIS treatment was demonstrated by contrasting the CTP relic densities numerically with those obtained using the most common RIS-subtraction schemes for two example models. For scalar singlet DM produced via $b\bar{b}$-annihilation close to the Higgs resonance, the narrow Higgs width renders the CTP, PVS, SDF and \texttt{micrOMEGAs} results practically indistinguishable, while adding decay and annihilation collision terms without any subtraction overestimates the relic density by about $\SI{75}{\percent}$, the real-part subtraction still by about $\SI{37}{\percent}$, and omitting the decay collision term underestimates it by about $\SI{25}{\percent}$. For vector-mediated scalar DM, the Hermitian self-energy already changes the relic density by about $\SI{15}{\percent}$ for $\Gamma_{Z'}^{\rm vac}/m_{Z'}\simeq10^{-3}$, the PVS scheme can even yield unphysical negative relic densities for large widths, and the \texttt{micrOMEGAs} result, which is based on a momentum-independent on-shell width, departs increasingly from the CTP result obtained with $\Pi^{\mathcal H}_{\mathrm T,\mathrm L}=0$ for $\Gamma_{Z'}^{\rm vac}/m_{Z'}\gtrsim0.04$. In both models, the Hard Thermal Loop approximation of the Hermitian self-energy reproduces the full one-loop result at the percent level up to moderately broad resonances, with deviations of up to $\SI{2.4}{\percent}$ and $\SI{6.5}{\percent}$ reached only at the largest widths considered.

Lastly, we would like to comment on the computational cost in terms of CPU time of the different approaches. While the runtime is strongly model and parameter dependent, we find that the numerical evaluation of the production rate in terms of the ``spectral kernels'' defined in Sec.~\ref{sec: Numerical Examples} is roughly equally expensive for all kernels given that the HTL form of the Hermitian mediator self-energy is used. To consistently treat the real intermediate state and to arrive at percent-level accurate relic abundance predictions, we therefore recommend to use the CTP production rate along with the HTL approximation for the Hermitian part of the mediator self-energy, while keeping the full momentum and mass dependence of the anti-Hermitian part.

We will soon generalize our framework to scenarios with multiple, possibly interfering, resonant mediators where RIS-subtraction schemes formulated for a single squared Breit--Wigner propagator are no longer directly applicable.

\acknowledgments

We thank J.R.\ Bade for his contributions in the early stages of this work, Vishnu P.K.\ for helpful discussions, M. Drewes for useful comments on the manuscript, and T.\ Bringmann and A.\ Pukhov for correspondence. Work in Münster was supported by the DFG through grant No. KL 1266/11-1. L.P.W. acknowledges support by Fonds de la Recherche Scientifique (FRS-FNRS). This work was made possible by Institut Pascal at Université Paris-Saclay with the support of the program “Investissements d’avenir” ANR-11-IDEX-0003-01, the P2I axis of the Graduate School of Physics of Université Paris-Saclay, as well as the CNRS-IRP project UCMN, IJCLab, CEA and ACME. The figures and Feynman diagrams presented in this paper were generated with \texttt{MatPlotLib} \cite{Hunter:2007} and \texttt{TikZ-Feynman} \cite{Ellis:2016jkw}.

\appendix

\section{Two-Point Functions on the CTP}\label{appendix: ctp props}
\subsection{Definitions and Relations} 

Here, we list the definitions of and relations between the CTP two-point functions $G$ that are used in this work, where $G$ may either stand for a propagator or a self-energy:
\begin{subequations}
    \begin{align}
    &G^R = G^{T} - G^{<} = G^> - G^{\bar T} &&\text{(Retarded)}\, , \\
    &G^A = G^{T} - G^{>} = G^{<} - G^{\bar T} && \text{(Advanced)}\, ,\\
    &G^\mathcal H=\frac{1}{2}(G^R+G^A) =  \frac{1}{2}(G^T-G^{\bar T}) &&\text{(Hermitian)}\, , \\
    &G^\mathcal A= \frac{i}{2}(G^R-G^A)=\frac{i}{2}(G^> - G^<)  &&\text{(Anti-Hermitian)}
\end{align}
\end{subequations}
with the Wightman functions $G^<= G^{12}$ and $G^>= G^{21}$ and the time-ordered two-point function $G^T =G^{11}$ and the anti-time-ordered two-point function $G^{\bar T} =G^{22}$. The anti-Hermitian component is also known as spectral function $\rho_G=2G^\mathcal A$.
In equilibrium, at temperature $T=1/\beta$, the Wightman functions obey the KMS relation that in momentum space becomes
\begin{align}
    G^>(p) = \pm e^{\beta p^0}G^<(p)
    \label{eq:KMS}
\end{align}
with $\pm$ for bosonic/fermionic field operators.

\subsection{Free Equilibrium Propagators}
The equilibrium free scalar propagators in momentum space, denoted by a subscript $0$, are 
    \begin{subequations}\label{eq: in equilibrium propagators scalar}
\begin{align}
      i\Delta^{11}_{{0}}(p)&=\frac{i}{p^2-m^2+i\epsilon}+ 2\pi \delta(p^2-m^2)f_\text{BE}(|p_0|)=\left( i\Delta_{{0}}^{22}(p) \right)^{\!*}\, ,\\
    i\Delta^{12}_{{0}}(p)&=2\pi \delta (p^2-m^2)[f_\text{BE}(|p_0|)+ \theta (-p_0)] \, , \\
    i\Delta^{21}_{{0}}(p)&=2\pi \delta (p^2-m^2)[f_\text{BE}(|p_0|)+ \theta (p_0)] \, .
\end{align}
\end{subequations}

Similarly, for fermions one has
\begin{subequations}\label{eq: fermionic equilibbirum propagators}
\begin{align}
i \slashed S_{{0}}^{11}(p) &= \frac{i(\slashed{p}+m)}{p^2 - m^2 + i\epsilon}- 2\pi (\slashed{p}+m)\,\delta(p^2-m^2)\,f_\text{FD}(|p_0|) = \gamma^0 \big(i \slashed S^{22}(p)\big)^\dagger \gamma^0 ,
\\i \slashed S_{{0}}^{12}(p) &= -2\pi (\slashed{p}+m)\,\delta(p^2-m^2)\left[f_\text{FD}(|p_0|)-\theta(-p_0)\right],
\\i \slashed S^{21}_{{0}}(p) &= -2\pi (\slashed{p}+m)\,\delta(p^2-m^2)\left[f_\text{FD}(|p_0|)-\theta(p_0)\right].
\end{align}
\end{subequations}
Here, $f_\text{BE}$ and $f_\text{FD}$ denote the Bose--Einstein and Fermi--Dirac distribution functions, respectively, for a vanishing chemical potential. 

For a massless gauge field at finite temperature that might appear as a mediator in our freeze-in context, unphysical degrees of freedom may arise in the thermal $\delta$-function part of the propagator due to gauge interactions. One strategy to avoid them is to generically include only the two physical vector-boson polarizations in this on-shell part.
This, however, requires the selection of a second reference direction besides the vector boson momentum. Since the presence of the thermal bath breaks Lorentz symmetry already, one simple choice is the four-velocity $u^\mu$ of the heat bath such that the polarization sum becomes
\begin{align}
    T^{\mu\nu} = -\eta^{\mu\nu} + \frac{k^\mu u^\nu + k^\nu u^\mu}{u \cdot k} - u^2 \frac{k^\mu k^\nu}{(u\cdot k)^2} \,,
\end{align}
which is nothing but the transverse projector $P_\mathrm T^{\mu\nu}$ from Eq.~\eqref{Eq: projectorsVectorMediator} in the frame where the plasma is at rest, $u^\mu = (1,\mathbf{0})$.
In the covariant Feynman gauge, the gauge boson propagator on the CTP then becomes~\cite{Landshoff:1992ne}
\begin{subequations}
\begin{align}
    i D^{11}_{0,\mu\nu}(p)&= - \frac{i\, \eta_{\mu\nu}}{p^2 + i\epsilon} + 2\pi\, T_{\mu\nu}\, \delta(p^2)\, f_\text{BE}(|p_0|) = \left( i D^{22}_{0,\mu\nu}(p) \right)^{\!*} \, ,\\
     i D^{12}_{0,\mu\nu}(p)&=2\pi\, T_{\mu\nu}\, \delta(p^2)\,[f_\text{BE}(|p_0|)+ \theta (-p_0)] \, , \\
     i D^{21}_{0,\mu\nu}(p)&=2\pi\, T_{\mu\nu}\, \delta(p^2)\,[f_\text{BE}(|p_0|)+ \theta (p_0)] \, .
\end{align}
\end{subequations}

In contrast, for a massive vector boson on the CTP~\cite{Evans:1993qg}, one can simply work in unitary gauge such that the corresponding propagator $D_{0,\mu\nu}^{ab}=  P_{\mu\nu}(p)\,\Delta_{{0}}^{ab}$ is the product of the scalar propagator and the polarization sum 
\begin{align}
    P^{\mu\nu}(p)  =- \eta^{\mu\nu} +\frac{p^\mu p^\nu}{m^2}\, .
\end{align}

One can see that, in equilibrium and in line with the Kadanoff--Baym ansatz, the free Wightman functions factorize into the free spectral density and the corresponding equilibrium distribution functions:
\begin{subequations}\label{eqs: Wightman KMS}
\begin{align}
    i\Delta_{{0}}^<(p) &= 2\Delta_{{0}}^\mathcal{A}(p)\,f_\text{BE}(p_0),
    &
    i\slashed S_{{0}}^<(p) &= -2\slashed S_{{0}}^\mathcal{A}(p)\,f_\text{FD}(p_0),\\
    i\Delta_{{0}}^>(p) &= 2\Delta_{{0}}^\mathcal{A}(p)\left(1+f_\text{BE}(p_0)\right),
    &
    i\slashed S_{{0}}^>(p) &= 2\slashed S_{{0}}^\mathcal{A}(p)\left(1-f_\text{FD}(p_0)\right).
\end{align}
\end{subequations}
with
\begin{subequations} \label{app: eqs: zero width forms}
\begin{align} \label{app:eq: spectral density equilibirum}
    \Delta_{{0}}^\mathcal A(p)&=\pi \operatorname{sgn}(p_0)\delta(p^2-m^2),
    \\
    \slashed S_{{0}}^\mathcal A(p)&=\pi \operatorname{sgn}(p_0)(\slashed p + m)\delta(p^2-m^2)
\end{align}
\end{subequations}
and similarly for vector fields.

\subsection{1PI-Resummed Scalar Equilibrium Propagators}
\label{sec:app:ctp-resummed-propagator}

In equilibrium, as a result of translation invariance, only the leading term of the gradient expansion survives, and the Dyson--Schwinger Eq.~\eqref{eq:DS} for the retarded and advanced propagators reduces in momentum space to
\begin{align} \label{eq:dyson schwinger in momentum space}
    \bigl(\Delta^{R/A}(p)\bigr)^{-1} = p^2 - m^2 - \Pi^{R/A}(p) \, .
\end{align}
At the level of the Dyson--Schwinger equation, the retarded and advanced propagators are thus determined by their own self-energies alone.
To obtain an expression for the 1PI-resummed Wightman function $\Delta^<$ required for the DM production rate in, e.g., Eq.~\eqref{eq:ctp-collision-general}, one can take the difference of the propagators in Eq.~\eqref{eq:dyson schwinger in momentum space},
\begin{align} \nonumber \Delta^R-\Delta^A = \Delta^R \left[ (\Delta^A)^{-1}-(\Delta^R)^{-1} \right] \Delta^A  = \Delta^R \left( \Pi^R-\Pi^A \right) \Delta^A
\end{align}
and apply the KMS relation Eq.~\eqref{eq:KMS}, that equivalently can be written as 
\begin{align} \Delta^<(p) &= f_\text{BE}(p_0) \left[ \Delta^R(p)-\Delta^A(p) \right], \qquad \Pi^<(p) = f_\text{BE}(p_0) \left[ \Pi^R(p)-\Pi^A(p) \right] \, .
\end{align} 
The final expression then reads
\begin{align}
\Delta^<(p) = \Delta^R(p)\Pi^<(p)\Delta^A(p). 
\end{align}
From there, the spectral density is
\begin{align}
\rho(p)=\frac{2\Pi^\mathcal A(p)}{\left[p^2-m^2-\Pi^\mathcal H(p)\right]^2+\left[\Pi^\mathcal A(p)\right]^2} .
\label{eq: app: resummed single mediator spectral density}
\end{align}

\section{Hermitian Mediator Self-energies}
\label{sec: app: hermitian self-energys}

The absorptive self-energies entering the mediator spectral densities are discussed in the main text. In this appendix, we collect the thermal Hermitian self-energies used in the numerical analysis. The vacuum contributions are understood to be absorbed into the renormalized mediator masses.

\subsection{Scalar Mediator}

We consider a scalar mediator $h$ coupled to a Dirac fermion $f$ through
\begin{align}
    \mathcal L_{\rm int} = -\frac{y_f}{\sqrt 2 } h \bar f f.
\end{align}
The Hermitian part of the scalar self-energy may be calculated from $\Pi_h^{\mathcal H}(p)=\operatorname{Re}\Pi_h^{11}(p)$. Inserting the free fermionic propagators, given in Eq.~\eqref{eq: fermionic equilibbirum propagators}, and using
\begin{align}
    \frac{i}{k^2-m_f^2+i\epsilon}=i\,\mathcal P\frac{1}{k^2-m_f^2}+\pi\delta(k^2-m_f^2),
\end{align}
the finite-temperature contribution can be written as 
\begin{align}
\Pi^\mathcal H_h(p, T)= & \sum_f \frac{y_f^2}{2}N_c^f\int \frac{\dd[4]k}{(2\pi)^3} \operatorname{Tr}[{(\slashed k + m_f) (\slashed k - \slashed p +m_f)}]\\\times& \Big[f_\text{FD}(|k_0|)\delta(k^2-m_f^2)\mathcal{P}\frac{1}{(k-p)^2-m_f^2}\hspace{-0.1em}+\hspace{-0.1em}f_\text{FD}(|k_0-p_0|)\delta((k-p)^2-m_f^2)\mathcal P \frac{1}{k^2-m_f^2}\Big]. \nonumber 
\end{align}
Since both the integration measure and the integrand are invariant under the shift $k\to p-k$, the two terms can be combined. Evaluating the trace and imposing the on-shell condition $k^2=m_f^2$, this becomes
\begin{align}
\Pi^\mathcal H_h(p, T)\hspace{-0.15em}=\hspace{-0.15em}\sum_f {y_f^2N_c^f}\int\hspace{-0.15em}\frac{\dd[4]k}{(2\pi)^3}\,4[-k\cdot p+2m_f^2]\,f_\text{FD}(|k_0|)\delta(k^2-m_f^2)\mathcal{P}\hspace{-0.1em}\frac{1}{(k-p)^2-m_f^2}.
\end{align}
To evaluate the integral systematically, it is convenient to decompose the expression into a scalar and a vector integral,
\begin{subequations}
\begin{align}
I(p, T)&=\int \frac{\dd[4]k}{(2\pi)^3}\,f_\text{FD}(|k_0|)\delta(k^2-m_f^2)\,\mathcal P\frac{1}{(k-p)^2-m_f^2},
\\
I^\mu(p, T)&=\int \frac{\dd[4]k}{(2\pi)^3}\,k^\mu f_\text{FD}(|k_0|)\delta(k^2-m_f^2)\,\mathcal P\frac{1}{(k-p)^2-m_f^2}.
\end{align}
\end{subequations}
The scalar integral can be reduced analytically to
\begin{align}
    I(p,T)=\frac{1}{4|\mathbf p|}\int_{m_f}^\infty \frac{\dd\omega_k}{(2\pi)^2}f_\text{FD}(\omega_k)\ln{\left|\frac{(p^2+2|\mathbf k||\mathbf p|)^2-4\omega_k^2p_0^2}{(p^2-2|\mathbf k||\mathbf p|)^2-4\omega_k^2p_0^2}\right|},
\end{align}
where $\omega_k=\sqrt{|\mathbf k|^2+m_f^2}$. The absolute value arises from taking the principal value and becomes relevant whenever the pole lies within the integration domain. The zeroth component of the vector integral is given by
\begin{align}
     I^0(p,T)&=\frac{1}{4|\mathbf p|}\int_{m_f}^\infty \frac{\dd\omega_k}{(2\pi)^2}\omega_kf_\text{FD}(\omega_k)\ln{\left|\frac{p^4-4(\omega_kp_0-|\mathbf k||\mathbf p|)^2}{p^4-4(\omega_kp_0+|\mathbf k||\mathbf p|)^2}\right|},
\end{align}
while the spatial components are
\begin{align}
  \mathbf I(p,T) &= \frac{\mathbf p}{32 \pi^2|\mathbf p|^3}\int_{m_f}^\infty{\dd\omega_k} f_\text{FD}(\omega_k)\\ \nonumber &\times\Big(8|\mathbf k| |\mathbf p|+ p^2\ln\Big|\frac{(p^2-2|\mathbf p||\mathbf k|)^2-4\omega_k^2p_0^2}{(p^2+2|\mathbf p||\mathbf k|)^2-4\omega_k^2p_0^2}\Big| -2\omega_k p_0 \ln\Big|\frac{p^4-4(p_0 \omega_k +|\mathbf k||\mathbf p| )^2}{p^4-4(p_0 \omega_k -|\mathbf k||\mathbf p| )^2}\Big|\Big).
\end{align}
In total, the Hermitian part of the self-energy is therefore given by
\begin{align}
 \nonumber    \Pi_h^\mathcal H(p,T)\hspace{-0.15em}=\hspace{-0.2em}\sum_f&\frac{y_f^2}{2} \frac{N_c^f}{|\mathbf p|} \hspace{-0.2em}\int_{m_{f}}^\infty \hspace{-0.2em}\frac{\dd\omega_k}{4\pi^2}\\ \times&\Bigg[{(4m_{f}^2-p^2)}\ln{\left|\frac{(p^2+2|\mathbf k||\mathbf p|)^2-4\omega_k^2p_0^2}{(p^2-2|\mathbf k||\mathbf p|)^2-4\omega_k^2p_0^2}\right|}+8|\mathbf k| |\mathbf p|\Bigg]f_\text{FD}(\omega_k).
\end{align}

\begin{figure}[t]
    \centering
    \includegraphics[width=0.5\linewidth]{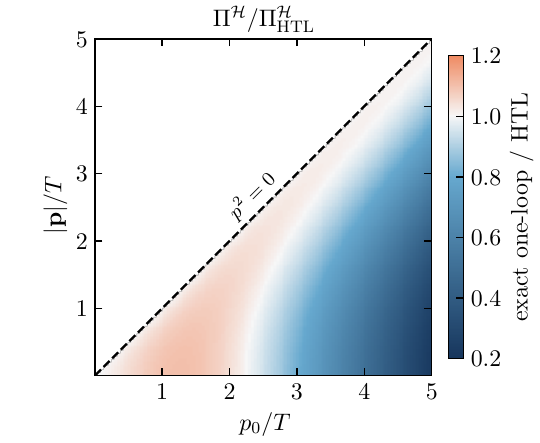}
    \caption{Heatmap of the ratio between the full one-loop Hermitian self-energy and its HTL approximation in a scalar Higgs-portal model. Only future-directed timelike momenta, $p^0>0$ and $p^2>0$, are shown, corresponding to the kinematic region relevant for DM production through the mediator. The dashed line indicates the light cone, $p^2=0$. We choose $T=\SI{100}{\giga\electronvolt}$, which lies in the production-dominated temperature range for the Higgs-portal benchmarks considered. Only the bottom-quark contribution is included, using $m_b=\SI{3}{\giga\electronvolt}$, approximately corresponding to the running bottom-quark mass at the energy scales examined. }
    \label{fig: hermitian self-energy higgsportals}
\end{figure}

The Hermitian self-energy may be evaluated in the Hard Thermal Loop approximation \cite{Pisarski:1988vd,BRAATEN1990569}. This approximation relies on a separation between hard scales of order $\mathcal{O}(\pi T)$ and soft scales of order $\mathcal{O}(gT)$. The external momentum $p$ is taken to be soft and is therefore small compared to the loop momentum $k$, whose hard contribution dominates the integral. In addition, the particle masses are assumed to be small compared to the temperature and can be neglected at leading order. The integrand can then be expanded in powers of $p/k\ll1$.
In the scenario considered here, one obtains
\begin{align}
   \Pi_{h,\text{HTL}}^\mathcal H(T)&=\sum_f y_f^2 N_c^f  \frac{T^2}{12},
\end{align}
which exhibits the characteristic $T^2$ scaling, in agreement with dimensional analysis. Figure \ref{fig: hermitian self-energy higgsportals} shows a comparison between the full one-loop calculation and the HTL approximation. As can be seen, the results agree well not only for $p_0\ll T$ and $|\mathbf p|\ll T$, but also for momenta near the light cone.

\subsection{Vector Mediator}
For the vector mediator model introduced in Sec.~\ref{sec:vetor-model}, the Hermitian self-energy is generated by quark loops. Adapting the finite-temperature one-loop results of Ref.~\cite{Drewes:2024wbw-Neff3} to our scenario, the transverse and longitudinal components are given by
\begin{align}
\Pi_\mathrm T^{\mathcal{H}}(p,T) &= \sum_q \frac{N_c^q g_q^2}{4\pi^2|\mathbf{p}|^3} \int_{m_q}^{\infty} \dd\omega\, f_\text{FD}(\omega) \Bigg[ 4|\mathbf{k}||\mathbf{p}| \left(p_0^2 + |\mathbf{p}|^2\right) - 2\ell_2(p,\omega)\, p_0 \,s\, \omega \\ \nonumber&\quad\qquad \qquad \qquad\qquad\qquad \quad \quad-\ell_1(p,\omega) \left( |\mathbf{p}|^2 s + 2p_0^2\omega^2 - 2|\mathbf{k}|^2|\mathbf{p}|^2 + \frac{s^2}{2} \right) \Bigg], \\
\Pi_\mathrm L^{\mathcal{H}}(p,T) &= \sum_q \frac{N_c^q g_q^2\, s}{4\pi^2|\mathbf{p}|^3} \int_{m_q}^{\infty} \dd\omega\, f_\text{FD}(\omega) \Big[ \ell_1(p,\omega)\left(s + 4\omega^2\right) + 4\ell_2 (p,\omega)\,p_0\,\omega - 8|\mathbf{k}||\mathbf{p}| \Big],
\end{align}
with $s=p^2$ and $|\mathbf k|=\sqrt{\omega^2-m_q^2}$. The logarithmic functions are defined as
\begin{align}
\ell_1(p,\omega) &= \ln\left| \frac{\left(s + 2|\mathbf{k}||\mathbf{p}|\right)^2 - 4p_0^2\omega^2}{\left(s - 2|\mathbf{k}||\mathbf{p}|\right)^2 - 4p_0^2\omega^2} \right|, \qquad
\ell_2(p,\omega) = \ln\left| \frac{s^2 - 4\left(p_0\omega + |\mathbf{p}||\mathbf{k}|\right)^2}{s^2 - 4\left(p_0\omega - |\mathbf{p}||\mathbf{k}|\right)^2} \right|.
\end{align}

\begin{figure}[t]
    \centering
    \includegraphics[width=0.95\linewidth]{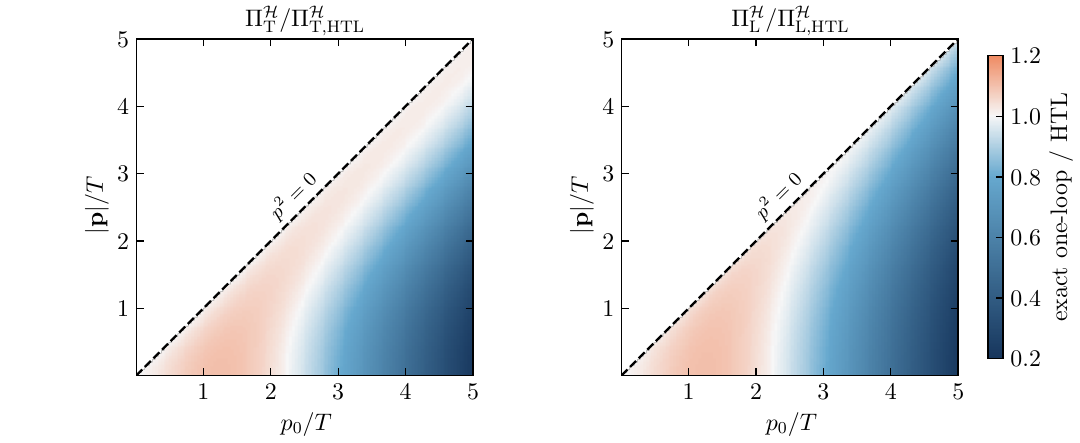}
    \caption{Heatmaps of the ratios $\Pi_ \mathrm T^\mathcal H/\Pi^\mathcal H_{\mathrm T\rm ,HTL}$ (left) and $\Pi_\mathrm L^\mathcal H/\Pi^\mathcal H_{\mathrm L,\rm HTL}$ (right) in the vector-mediator model. Only future-directed timelike momenta, $p^0>0$ and $p^2>0$, are shown, corresponding to the kinematic region relevant for DM production through the mediator. The dashed line indicates the light cone, $p^2=0$. We choose $T=\SI{5000}{\giga\electronvolt}$, which lies in the production-dominated region for the parameters considered. Since this temperature is well above the electroweak scale, all six quark flavors are treated as massless in this plot. }
    \label{fig: Vectormediator hermitian self-energys}
\end{figure}
The respective HTL expressions are given in Eq.~\eqref{eq:HTL vector}.
Their accuracy is illustrated in Fig.~\ref{fig: Vectormediator hermitian self-energys}, which shows the ratios $\Pi_\mathrm T^\mathcal H/\Pi^\mathcal H_{\mathrm T\rm ,HTL}$  and $\Pi_\mathrm L^\mathcal H/\Pi^\mathcal H_{\mathrm L,\rm HTL}$. The HTL approximation agrees well with the full one-loop result for momenta close to the light cone.

\bibliographystyle{JHEP}
\bibliography{paper}

\end{document}